\documentclass{aa}  
\usepackage{natbib}
\bibpunct{(}{)}{;}{a}{}{,} 
\usepackage{graphicx}
\usepackage[varg]{txfonts}
\usepackage{hyperref}
\hypersetup{colorlinks, allcolors = {blue}}
\begin{document} 
    
    \title{Impacts of nonthermal emission on the images of a black hole shadow and extended jets in two-temperature GRMHD simulations}
    \titlerunning{Impact of nonthermal emission on the images of a black hole shadow and jets}

    \author{Mingyuan Zhang\inst{1,2}
        \and
            Yosuke Mizuno\inst{1,3,4}
        \and
            Christian M. Fromm\inst{5,4,6}
        \and
            Ziri Younsi\inst{7}
        \and
            Alejandro Cruz-Osorio\inst{8}
            }
    \authorrunning{Zhang et al.}

    \institute{Tsung-Dao Lee Institute, Shanghai Jiao Tong University, Shanghai, 201210, People's Republic of China\\
            \email{mzhang22@sjtu.edu.cn, mizuno@sjtu.edu.cn}
        \and
             College of Physics, Jilin University, Changchun, 130012, People's Republic of China
        \and
            School of Physics and Astronomy, Shanghai Jiao Tong University, Shanghai, 200240, People’s Republic of China
        \and
            Institut für Theoretische Physik, Goethe Universität, Max-von-Laue-Str. 1, D-60438 Frankfurt, Germany
        \and
            Institut f\"ur Theoretische Physik und Astrophysik, Universit\"at W\"urzburg, Emil-Fischer-Str. 31, D-97074 W\"urzburg, Germany
        \and
            Max-Planck-Institut f\"ur Radioastronomie, Auf dem H\"ugel 69, D-53121 Bonn, Germany
        \and
            Mullard Space Science Laboratory, University College London, Holmbury St. Mary, Dorking, Surrey, RH5 6NT, UK
        \and
            Instituto de Astronom\'{\i}a, Universidad Nacional Aut\'onoma de M\'exico, AP 70-264, Ciudad de M\'exico 04510, M\'exico
            }


 
  \abstract
   {The recent 230\,GHz observations from the Event Horizon Telescope (EHT) collaboration are able to image the innermost structure of the M87 galaxy showing the shadow of the black hole, a photon ring, and a ring-like structure that agrees with thermal synchrotron emission from the accretion disc. However, at lower frequencies, M87 is characterized by a large-scale jet with clear signatures of nonthermal emission. {It is necessary to explore the impacts of nonthermal emission on black hole shadow images and extended jets, especially at lower frequencies.}} 
   {In this study, we {aim to compare models with different electron heating prescriptions to one another and to investigate how these prescriptions and nonthermal electron distributions may affect black hole shadow images and the broadband spectrum energy distribution (SED) function.}}
   {We performed general relativistic radiative transfer (GRRT) calculations in various two-temperature general relativistic magnetohydrodynamic (GRMHD) models utilizing different black hole spins and different electron heating prescriptions coupled with different electron distribution functions (eDFs).} 
   {Through a comparison with GRRT images and SEDs, we found that when considering a variable $\kappa$ eDF, the parameterized prescription of the $R-\beta$ electron temperature model with $R_{\mathrm{h}}$ = 1 is similar to the model with electron heating in the morphology of images, and the SEDs at a high frequency. This is consistent with previous studies using the thermal eDF. However, the nuance between them could be differentiated through the diffuse extended structure seen in GRRT images, especially at a lower frequency, and the behavior of SEDs at low frequency. The emission from the nearside jet region is enhanced in the case of electron heating provided by magnetic reconnection and it will increase if the contribution from the regions with stronger magnetization is included or if the magnetic energy contribution to $\kappa$ eDF mainly in the magnetized regions is considered. Compared with the thermal eDF, the peaks of the SEDs shift to a lower frequency when we consider nonthermal eDF.}
   {}


   \maketitle
%



\section{Introduction}
\label{introduction}

The supermassive black hole M87\,$^{*}$ is an ideal object that can be observed directly, and the study of its black hole shadow is of great scientific significance. 
The Event Horizon Telescope (EHT) collaboration utilized very-long-baseline interferometer (VLBI) technology to network eight radio telescopes around the world into forming a virtual telescope with an effective diameter equivalent to the diameter of the Earth, which makes the direct observation of supermassive black holes possible. 
There are two main targets: the black hole at the Galactic center, Sgr~A$^{*}$, and the black hole in the nearby elliptical galaxy M87. 
M87 is somehow an ideal object for direct observation due to its characteristics of long periods and a lack of interstellar scattering interference \citep{2006ApJ...648L.127B} with the presence of a powerful radio jet \citep{2009ApJ...697.1164B}.
Theoretically, when one looks at the black hole, one can see the black hole shadow, which is the silhouette of the unstable light region around the event horizon \citep[][]{1973ApJ...183..237C, 1994NuPhB.423..532D, 2014PhRvD..89l4004G, 2016PhRvD..94h4025Y}.
The study of black hole shadows provides insights into their mass, inclination angle, and spin, making it one of the most rigorous tests of general relativity.

Numerical modeling helps us understand the physics behind black hole shadows.
Observations provide a reference for model prediction, while the synthetic images are made by the general relativistic magnetohydrodynamic (GRMHD) simulation and general relativistic radiative transfer (GRRT) calculations. Comparison between synthetic images and observations shows whether the physical models are correct.
At present, numerical modeling has been effective. 
A ring-like structure as the most direct observational evidence of a black hole shadow is mostly consistent with the synthetic models given by the GRMHD simulations \citep{2019ApJ...875L...1E, 2019ApJ...875L...2E, 2019ApJ...875L...3E, 2019ApJ...875L...4E, 2019ApJ...875L...5E, 2019ApJ...875L...6E}.

General relativistic magnetohydrodynamic simulations are helpful in improving our understanding of accretion physics.
Generally, M87 is believed to have a low accretion rate consisting of a geometrically thick but optically thin accretion disk in a state of radiatively inefficient accretion flow (RIAF) \citep[][]{1995ApJ...444..231N, 2014ARA&A..52..529Y}.
It has been found that the mass accretion rate of M87 is several orders of magnitude smaller than the Eddington limit, and its brightness is much smaller than the corresponding Eddington brightness  \citep[][]{2009ApJ...699..626H}. Similarly, Faraday rotation measurements provide indirect evidence for the low accretion rate of M87 \citep[][]{2003ApJ...588..331B, 2007ApJ...654L..57M, 2014ApJ...783L..33K}. 
In order to study the radiative signature at the event horizon scale in M87, many GRMHD simulations have been performed in the RIAF state around the rotating black hole \citep[][]{1999MNRAS.303..343K, 2003ApJ...589..458D, 2007CQGra..24S.259N, 2009ApJ...706..497M, 2014A&A...570A...7M, 2016A&A...586A..38M, 2010ApJ...717.1092D, 2012ApJ...755..133S, 2018A&A...612A..34D, 2019A&A...632A...2D, 2018JPhCS1031a2008O, 2018NatAs...2..585M}. 

General relativistic radiative transfer calculations utilize GRMHD simulation data to compute black hole shadow images, light curves, and spectrum energy distributions (SEDs). GRMHD codes typically focus only on ions that are important for fluid evolution and provide little information about electrons. Therefore, the calculation of important parameters for modeling the electromagnetic radiation of accreting black holes, such as the electron distribution function (eDF) and electron temperature, becomes an open problem \citep{2019A&A...632A...2D}. In terms of the {eDF}, in the GRMHD simulation of M87, the EHT collaboration assumed that the {eDF} of all simulation regions is the Maxwell-J$\Ddot{\mathrm{u}}$ttner distribution \citep{2019ApJ...875L...5E}. However, the energy distribution of electrons depends on energy dissipation, particle acceleration, and thermalization \citep{2014ARA&A..52..529Y}. 
If only the thermal distribution is considered, some potentially important nonthermal effects will be ignored, such as magnetic reconnection and turbulent dissipation, which will accelerate electrons to the nonthermal power law distribution \citep[][]{2010ApJ...708.1545D, 2013ApJ...773..118H}. In addition, some features of M87 caused by electron acceleration have been observed in the near-infrared and optical bands \citep{2016MNRAS.457.3801P}. 
{Therefore, the impacts of nonthermal distribution merit consideration.} In view of the influence of nonthermal distributed electrons on images, a so-called $\kappa$ distribution \citep{1968JGR....73.2839V}, which can both reproduce thermal-electron distribution and present power law distribution characteristics at high energies by adjusting parameters, attracts people's interest.
Later, one can use the GRMHD simulations in a standard accretion and normal evolution (SANE) state \citep{2012MNRAS.426.3241N, 2013MNRAS.436.3856S} in GRRT calculations to study the black hole shadow using the $\kappa$ {eDF} \citep[][]{2018A&A...612A..34D, 2019A&A...632A...2D}. 
Recently, the application of the $\kappa$ distribution to magnetically arrested disk (MAD) models \citep{2003PASJ...55L..69N, 2011MNRAS.418L..79T} has reproduced the wide opening angle jet morphology at 86~GHz and fit the broadband spectrum of M87 from the radio to the near-infrared bands \citep{2022NatAs...6..103C}. Besides, compared with SANE models, \cite{2022A&A...660A.107F} indicate that the broad and highly magnetized jet illuminated by the nonthermal electrons stemming from the $\kappa$ distribution emerges in $R-\beta$ models with the MAD state. Furthermore, \cite{2023ApJ...959L...3D} show the propagation of waves along the shear layer of the jet wind using the $\kappa$ distribution in the MAD regime.

As is shown in {VLBI multi-wavelength observations of M87\,$^{*}$ \citep{2021ApJ...911L..11E}, the spectral index, $\alpha$, is estimated between $-1 \leq \alpha \leq 0$, which indicates the existence of nonthermal distributions of electrons. Motivated by the indication of the existence of electrons in observation, to} self-consistently understand the contribution of the nonthermal emission in the images, it is necessary to investigate the impacts of nonthermal emission on black hole shadow images in two-temperature GRMHD simulations.
A previous study using a three-dimensional two-temperature GRMHD simulation {with thermal distribution} on the MAD state of M87 has been conducted by \cite{2021MNRAS.506..741M}. 
The results of the elementary parameterized $R-\beta$ model were found to be consistent with those of complicated electron-heating models at 230~GHz when considering the Maxwell-J$\Ddot{\mathrm{u}}$ttner distribution.
However, the impacts of nonthermal electrons on shadow images and the features at different frequencies were not investigated in the previous work. 
Therefore, the nonthermal distribution, or the $\kappa$ distribution, is left for this paper to study.

This paper is arranged as follows: Section \ref{introduction} gives a brief overview of the background. In Sect. \ref{setup}, we introduce two electron heating prescriptions used in GRMHD simulations and the nonthermal eDFs used in the GRRT calculation. The results of the GRRT calculation, light curves, and SEDs are shown in Sect. \ref{results}. In Sect. \ref{discussion}, we discuss our results and the limitations of our study. 

{Throughout this paper, we adopt units where the speed of light is $c = 1$ and the gravitational constant is $G = 1$. We absorb a factor of $\sqrt{4\pi}$ into the definition of the magnetic field 4-vector, $b^{\mu}$.}
\section{Numerical setup}
\label{setup}

In this section, the computational methods are described. This includes the two-temperature GRMHD simulation data and two electron heating prescriptions used in this paper. Subsequently, an account is given of the traditional methods and the new methodology for the {eDFs} used in the GRRT calculation at the end of this section.

\subsection{General relativistic magnetohydrodynamic setup} 
\label{2.1}

We used the same GRMHD simulation data in previous studies \citep[][]{2021MNRAS.506..741M, 2022A&A...660A.107F}.
There are three-dimensional two-temperature GRMHD simulations of magnetized accretion flows using the BHAC code \citep{2017ComAC...4....1P, 2019A&A...629A..61O}. The metric adopted in the simulation is spherically modified Kerr-Schild (MKS) coordinates.
The simulations were performed up to $t=15\,000\,\mathrm{M}$. They reach a quasi-stationary MAD state {(see Appendix \ref{AppendixD} for more details)}. The time evolution of electron temperature in two-temperature GRMHD simulations is based on solving the electron entropy equation \citep{2015MNRAS.454.1848R,2021MNRAS.506..741M}. The detailed initial setup is described in the following papers \citep[][]{2021MNRAS.506..741M, 2022NatAs...6..103C, 2022A&A...660A.107F}.

In the GRMHD simulation, electron heating is provided by grid-scale dissipation. 
{Although it is mostly numerical, we consider grid-scale dissipation to mimic physical processes to heat the electrons \citep{2015MNRAS.454.1848R}.}
The physical processes include turbulent heating, magnetic reconnection, shock waves, and Ohmic dissipation. 
This paper uses two heating models: turbulence \citep{2019PNAS..116..771K} and magnetic reconnection \citep{2017ApJ...850...29R}.

\subsection{General relativistic radiative transfer setup}
In order to calculate images of black hole shadows from GRMHD simulations, this paper uses the GRRT code \texttt{BHOSS} \citep{2012A&A...545A..13Y,2020IAUS..342....9Y,Younsi2023}, which solves the equations of covariant radiative transfer via the ray-tracing method. 
The synchrotron radiation from electrons, as a radiation mechanism, is considered in this paper. 

This paper considers two forms of the {eDF}.
One is the Maxwell-J\"uttner distribution given by {Eq.}~\eqref{eq5} and the other is the $\kappa$ distribution, which can simultaneously represent thermal electron distribution properties and power law characteristics by adjusting certain parameters of {Eq.}~\eqref{eq6}. 
The Maxwell-J\"uttner distribution is expressed as follows:
\begin{equation} \label{eq5}
\frac{d n_{\mathrm{e}}}{d \gamma_{\mathrm{e}}}=\frac{n_{\mathrm{e}}}{\Theta_{\mathrm{e}}} \frac{\gamma_{\mathrm{e}} \sqrt{\gamma_{\mathrm{e}}^{2}-1}}{K_{2}\left(1 / \Theta_{\mathrm{e}}\right)} \exp \left(-\frac{\gamma_{\mathrm{e}}}{\Theta_{\mathrm{e}}}\right)\,,
\end{equation}
where $n_{\mathrm{e}}$ is the electron number density, $\gamma_{\mathrm{e}}$ is the electron Lorentz factor, $K_{2}$ is the Bessel function of the second kind, and $\Theta_{\mathrm{e}}$ is the dimensionless electron temperature \citep[see, e.g.,][]{2021MNRAS.506..741M}. 

The relativistic nonthermal $\kappa$ distribution \citep{2006PPCF...48..203X} is expressed as follows:
\begin{equation} \label{eq6}
\frac{d n_{\mathrm{e}}}{d \gamma_{\mathrm{e}}}=N \gamma_{\mathrm{e}} \sqrt{\gamma_{\mathrm{e}}^{2}-1}\left(1+\frac{\gamma_{\mathrm{e}}-1}{\kappa w}\right)^{-(\kappa+1)},
\end{equation}
where $N$ is the normalization factor \citep{2016ApJ...822...34P,2019A&A...632A...2D} and $\kappa$ is related to the slope of the power law distribution, $s=\kappa-1$.
When $\gamma_{\mathrm{e}}$ is large, particles satisfy $d n_{\mathrm{e}} / d \gamma_{\mathrm{e}} \propto \gamma_{\mathrm{e}}^{-s}$ and {the} nonthermal $\kappa$ distribution approximates the power law distribution.
The parameter $w$ specifies the width of a $\kappa$ distribution.
Considering the contribution of both thermal energy and magnetic energy to heating and accelerating electrons \citep{2019A&A...632A...2D, 2022NatAs...6..103C,2022A&A...660A.107F}, the specific expression of $w$ is written as follows:
\begin{equation} \label{eq4}
w:=\frac{\kappa-3}{\kappa} \Theta_{\mathrm{e}}+\frac{\varepsilon}{2}\left[1+\tanh \left(r-r_{\mathrm{inj}}\right)\right] \frac{\kappa-3}{6 \kappa} \frac{m_{\mathrm{p}}}{m_{\mathrm{e}}} \sigma\,,
\end{equation}
where $r_{\mathrm{inj}}$ is the injection radius, $m_{\mathrm{e}}$ is the electron mass, $m_{\mathrm{p}}$ is the proton mass, $\sigma=b^2/\rho$ is the magnetization, $b^2$ is the square of a 4-magnetic field, $\rho$ is the fluid rest-mass density, and $\varepsilon$ is a tunable parameter for the region with a radius larger than $r_{\mathrm{inj}}$. The energy is dominated by thermal energy with a limit of $\sigma \ll 1$, while the magnetic energy contributes to magnetized regions. We consider $\varepsilon$ to be zero or nonzero. We set $\varepsilon=0$ by default to study the impacts of the $\kappa$ distribution with thermal energy and show the results with a constant $\varepsilon=0.5$ in Appendix \ref{AppendixA} for the discussion of the magnetic energy contribution to {the GRRT images and the spatial distribution of $w$}. The jet stagnation surface is a potential injection site and defines the injection radius.
The stagnation surface is located at $u^{r} = 0$, where the potential injection radius is usually between $5~\mathrm{M}$ and $10~\mathrm{M}$ \citep{2018ApJ...868..146N}.
We therefore assumed $r_{\mathrm{inj}} = 10\,\mathrm{M}$ in this study.

Two $\kappa$ values were used throughout: a fixed $\kappa$ value with $\kappa = 3.5$, and a variable value of $\kappa$ defined to be parametrically dependent on magnetization, $\sigma$, and plasma beta, $\beta=p_{\mathrm{g}}/p_{\mathrm{m}}$, where $p_{\mathrm{g}}$ is the fluid pressure and $p_{\mathrm{m}}=b^2/2$ is the magnetic pressure: 
\begin{equation}
\kappa:=2.8+0.7 \sigma^{-1/2} + 3.7\,\sigma^{-0.19} \, \tanh \left(23.4\, \sigma^{0.26} \,\beta\right) \,,
\end{equation}
which was obtained empirically from particle-in-cell (PIC) simulations of the Harris current sheet \citep{2018ApJ...862...80B} (see also \citep{Meringolo2023} for results using PIC simulations of turbulent plasma). The distribution of accelerated ions is well fit by the fixed $\kappa$ distribution shown in the PIC simulation \citep{2016PhRvL.117w5101K}. We also considered a fixed $\kappa$ distribution {for electrons} and set $\kappa = 3.5$ based on the prediction of the spectral index of $\alpha \approx -0.7$ for MAD models regardless of spin \citep{2023MNRAS.519.4203R}, with $\alpha = -(\kappa - 2)/2$.

In this study, we used two methods to obtain the electron temperature for the GRRT calculations.
The first method estimated the electron temperature from a parametric prescription, here the ``$R-\beta$'' model \citep{2009ApJ...706..497M}.
The second method directly calculated the electron temperature from the two-temperature GRMHD simulations \citep{2021MNRAS.506..741M}.
The parametric $R-\beta$ model is given by
\begin{equation} \label{eq9}
\frac{T_{\mathrm{i}}}{T_{\mathrm{e}}}=\frac{R_{\mathrm{l}}+R_{\mathrm{h}}\,\beta^{2}}{1+\beta^{2}}\,,
\end{equation}
where $R_{\mathrm{l}}$ and $R_{\mathrm{h}}$ are hyperparameters, and $T_{\mathrm{i}}$ is the plasma ion temperature and electron temperature, $T_{\mathrm{e}}=m_{\mathrm{e}} c^{2} \Theta_{\mathrm{e}} / k_{\mathrm{B}}$, in c.g.s units, where $k_{\mathrm{B}}$ is Boltzmann's constant and $c$ is the speed of light. 
When $R_{\mathrm{l}}$ is fixed, the ratio of ion-to-electron temperatures in the strongly magnetized region ($\beta \ll 1$) is approximately $R_{\mathrm{l}}$, and thus the temperature of electrons in the highly magnetized jet does not change with $R_{\mathrm{h}}$.
Conversely, in weakly magnetized regions, the ratio of ion-to-electron temperatures is approximately $R_{\mathrm{h}}$.
We note that the image comparison results do not depend on the value of $R_{\mathrm l}$ \citep{2021MNRAS.506..741M}.
Therefore, we fixed $R_{\mathrm l}=1$ and varied $R_{\mathrm h}$ as $R_{\mathrm h}=(1,\, 10,\, 80,\, 160)$. 

For the GRRT calculation, we modeled M87\,$^{*}$ with a mass, $M_{\mathrm{BH}} = 6.5 \times 10^9 \, \mathrm{M_{\odot}}$, at a distance, $D = 16.8 \, \mathrm{Mpc}$ \citep{2019ApJ...875L...6E}.
The field of view (FoV) was initialized to be 320 $\mathrm{\mu as}$ in both directions and the resolution was $640 \times 640$ pixels.
The calculations were performed from $t = 14\,000\,\mathrm{M}$ to $t=15\,000\,\mathrm{M}$, with a $10\,\mathrm{M}$ cadence.
The time-averaged flux was set to be $0.5~{\mathrm{Jy}}$ at 230~GHz at $163^{\circ}$ in the entire current FoV; thus, at any given time the images at 86~GHz or 43~GHz, and at $163^{\circ}$ or $60^{\circ}$, share the same mass accretion rate as those calculated at 230~GHz at $163^{\circ}$. The ceiling to exclude regions with strong magnetization was $\sigma_{\mathrm{cut}}=1$ by default and we varied $\sigma_{\mathrm{cut}}$ as $\sigma_{\mathrm{cut}}=(1,\, 2,\, 5,\, 10)$ in Sect. \ref{3.6}. 

\section{Results}
\label{results}

\subsection{ General relativistic radiative transfer calculation}

\begin{figure*}
\centering
    \includegraphics[width=0.85\linewidth]{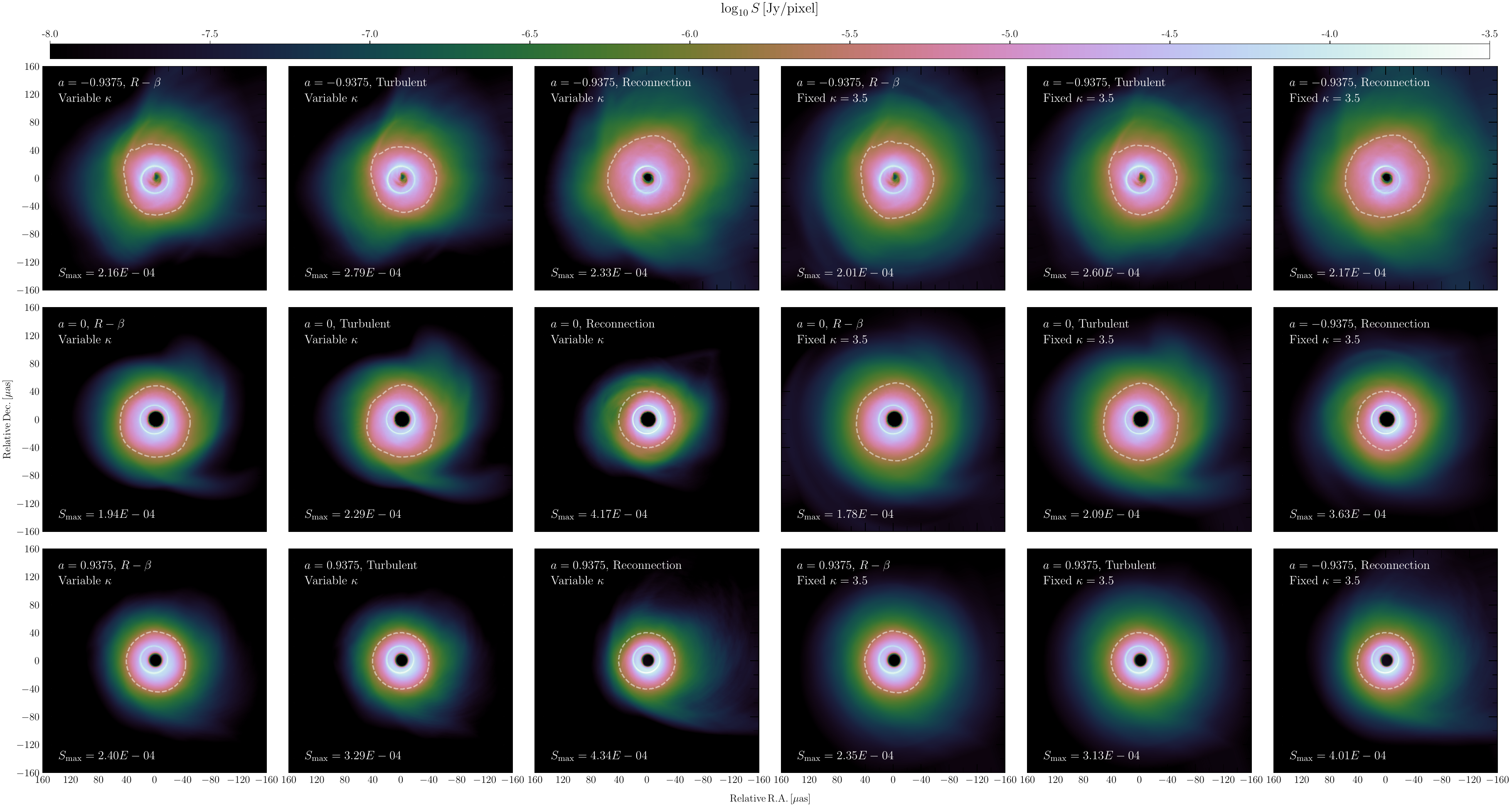}
    \caption{Time-averaged GRRT images of MAD simulations with various black hole spins and eDFs at 230~GHz with an inclination angle of $163^{\circ}$. 
    From top to bottom: Black hole spins with $a = -0.9375$, $0$, and $0.9375$, respectively.
    From left to right: The variable $\kappa$ eDF using the $R-\beta$ model with $R_{\mathrm{h}} = 1$, turbulent and reconnection heating, and a fixed $\kappa$ eDF, with $\kappa=3.5$ using the $R-\beta$ model with $R_{\mathrm{h}} = 1$, turbulent and reconnection heating prescriptions. 
    The dashed white contour indicates the curve with $1 \%$ of the maximum flux.
    The time-averaged total flux is $0.5$ Jy at 230~GHz, with a simulation time span from $t=14\,000$ to $15\,000\, \mathrm{M}$.} 
    \label{figure1}
\end{figure*}

\begin{figure*}
\centering
    \includegraphics[width=0.85\linewidth]{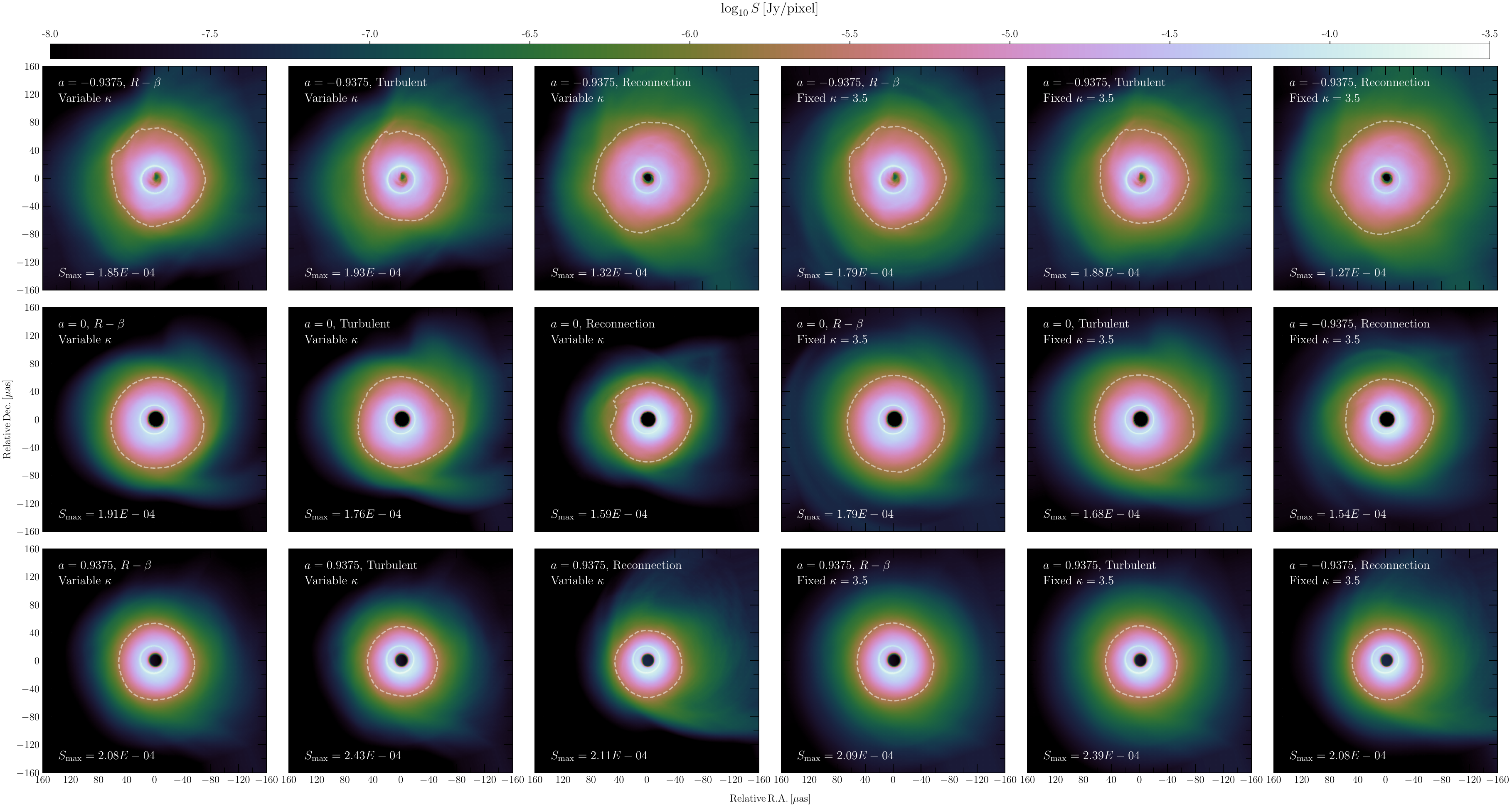}

    \caption{Same as Fig.~\ref{figure1} but shown in 86~GHz.}
    \label{figure2}
\end{figure*}

\begin{figure*}
\centering
    \includegraphics[width=0.85\linewidth]{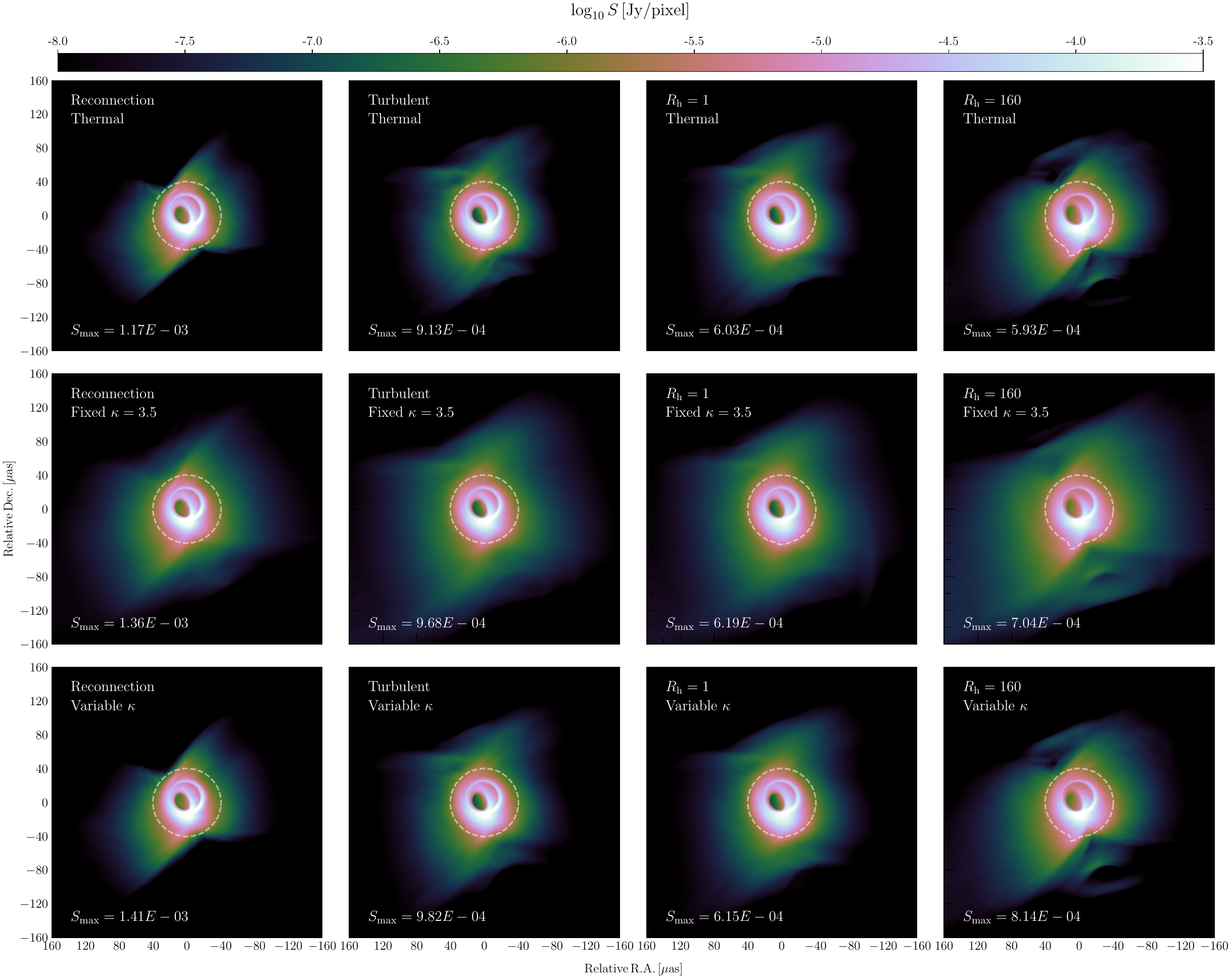}
    \caption{Time-averaged GRRT images of MAD simulations with a black hole spin of $a = 0.9375$ at 230~GHz and an inclination angle of $60^{\circ}$. 
    From top to bottom: The eDF models using thermal, fixed $\kappa=3.5$, and variable $\kappa$, respectively.
    From left to right: Electron heating prescriptions using reconnection heating, turbulent heating, and the $R-\beta$ model with $R_{\mathrm{h}} = 1$ and $R_{\mathrm{h}} = 160$, respectively.
    The dashed white contour indicates the curve with $1 \%$ of the maximum flux.
    The time-averaged total flux is $0.5~{\mathrm{Jy}}$ from $t=14\,000$ to $15\,000 \,\mathrm{M}$ at $163^{\circ}$. 
    }
    \label{figure3}
\end{figure*}

Figures~\ref{figure1} and \ref{figure2} depict time-averaged GRRT images at 230~GHz and 86~GHz using different nonthermal eDFs, spins, and electron heating prescriptions in logarithmic intensity scaling. 
In general, all cases show a bright photon ring with some diffuse extended {structures}. We see that the extended structure is more prominent in 86~GHz.


In the comparison between electron heating and parameterized $R-\beta$ prescriptions, the results indicate that the morphological structure of the turbulent heating model is matched with the one using the $R-\beta$ model with $R_{\mathrm{h}} = 1$ in all spin cases, using the variable $\kappa$ eDF. 

Compared to images of variable $\kappa$ eDF cases, extended emission is prominently displayed in the images using fixed $\kappa$ eDF, as was expected. This is because we employed a uniformly large nonthermal electron population, even in regions with low magnetization. 
It should be noted that the central bright ring is mainly attributed to the region near the black hole, mostly by thermal components \citep{2019ApJ...875L...5E}, while the extended diffuse emission comes from the jet sheath or wind region that is profoundly influenced by the choice of eDF.
When considering a model with fixed $\kappa=3.5$, compared with thermal or variable $\kappa$ eDF, the nonthermal power law tail of the eDF is well extended (see Fig. 4 in \cite{2022A&A...660A.107F}).
Therefore, in the fixed $\kappa$ eDF case, the extended diffuse emission becomes more luminous, while the maximum flux becomes slightly lower than that of the variable $\kappa$ eDF because we set the total flux to be fixed.

Moreover, as is shown in the corotating cases, there is a semi-arc-like structure in the images of the reconnection heating model.
The extended semi-arc-like structure represents the enhancement of emission on the nearside jet shown in Fig.~\ref{figure4}, in the reconnection heating cases.
Such a semi-arc-like extended structure is likely to be the propagation of Alfv\'en waves or the emerging magnetic flux tube \citep{Moriyama2023, 2023MNRAS.522.2307J}.

Conversely, the presence of extended structure not only relies on heating prescriptions but also relates to their spins.
The differences between the jet width and the jet length for the different spins can be related to the variability of the accretion flow and the accreted magnetic flux \citep{2022A&A...660A.107F}.
The extended structure in nonrotating cases would be different from the highly rotating cases due to the emission mainly coming from outflowing winds or accretion flows instead of highly magnetized powerful jets due to the lack of an ergosphere \citep{1977MNRAS.179..433B}. 
In counter-rotating cases, the rotation direction between the accretion flow and the black hole is opposed. 
Consequently, at the same inclination angle, the flow direction is different from the corotating cases, which causes the distinct enhancement of extended jet regions. 
In counter-rotating cases, the inner stable circular orbit (ISCO) position ($R_{\mathrm{ISCO}} \approx 8.8 \,\mathrm{M}$) is far from the black hole horizon, inducing a dim photon ring and enhancement of the extended structure. The counter-rotating cases exhibit more elongated or elliptical structures, while corotating cases show homogeneous or spherical structures. As is shown in Fig.~\ref{figure2} at 86~GHz, the differences stemming from spins become bigger.

Figure~\ref{figure3} presents the time-averaged corotating GRRT images at 230~GHz with the inclination angle $i=60^{\circ}$ on a logarithmic scale in different electron heating prescriptions and eDF models. 
In comparing the thermal and the $\kappa$ eDFs, images of fixed $\kappa=3.5$ have a clear diffuse extended structure and this trend is also seen in the images with an inclination angle of $163^{\circ}$. 
The images of variable $\kappa$ are analogous to thermal ones except for the small differences in the extended structure and the maximum flux. The images using reconnection heating lead to a smaller vertical extended structure, in contrast with the ones of the turbulent heating or the $R-\beta$ prescriptions. This difference in vertically extended structure results from the difference in the contribution to the emission from the midplane region (see Appendix \ref{AppendixB} for more details).  

\subsection{Image decomposition}
\begin{figure*}
\centering
    \includegraphics[width=0.85\linewidth]{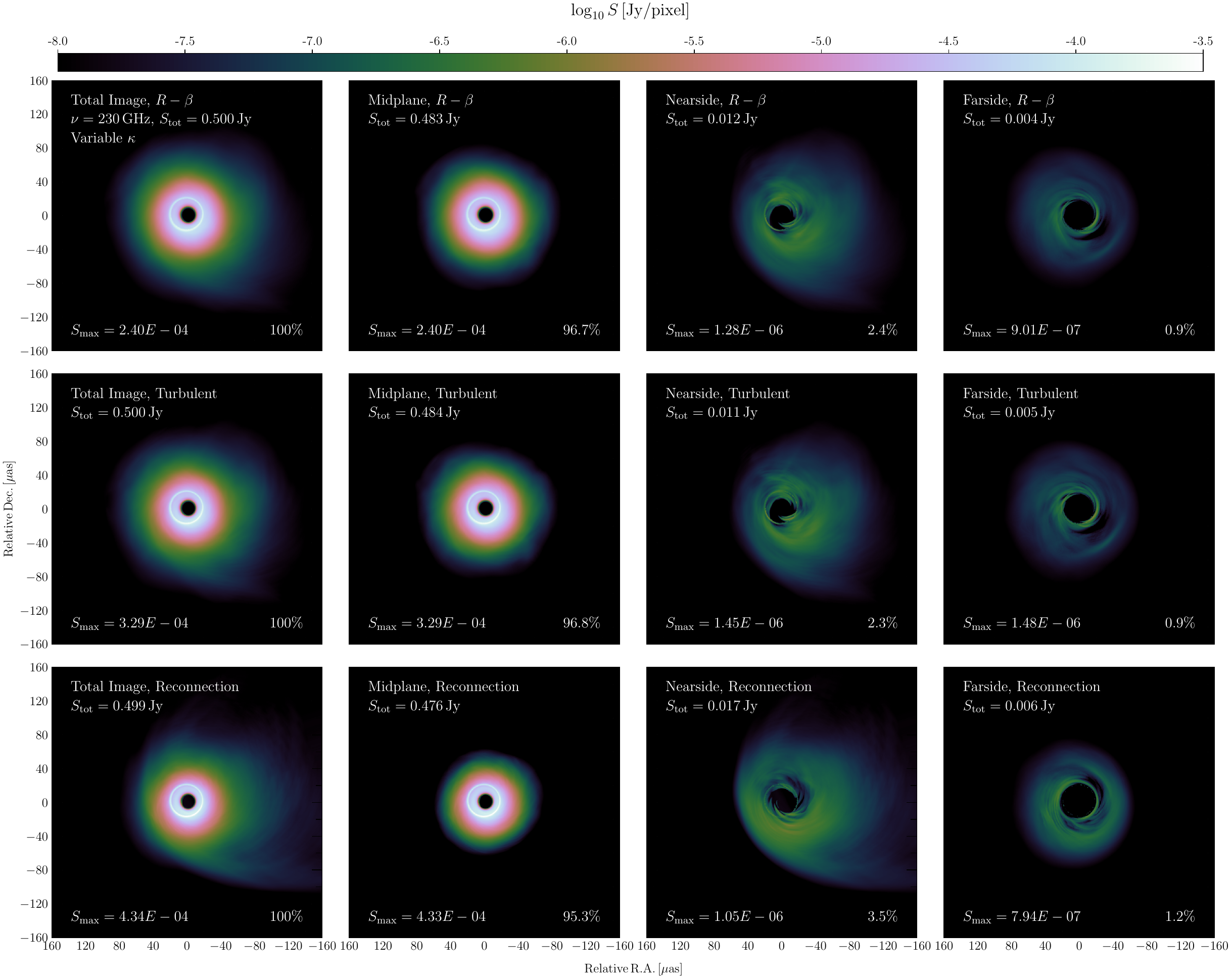}
    \caption{Time-averaged GRRT decomposed images with a black hole spin of $a = 0.9375$ at 230~GHz with a $163^{\circ}$ inclination angle. 
    From top to bottom: Images using the $R-\beta$ model with $R_{\mathrm{h}} = 1$, the turbulent heating model, and the reconnection heating model, respectively. 
    From left to right, the emissions come from every region: the whole region is depicted first, then the midplane, the nearside jet, and the farside jet, respectively.
    The eDF of all of the images is variable $\kappa$ and the time-averaged total flux is $0.5$ Jy at 230~GHz from $t=14\,000$ to $15\,000 \,\mathrm{M}$.}
    \label{figure4}
\end{figure*}
\begin{figure*}
\centering
    \includegraphics[width=0.85\linewidth]{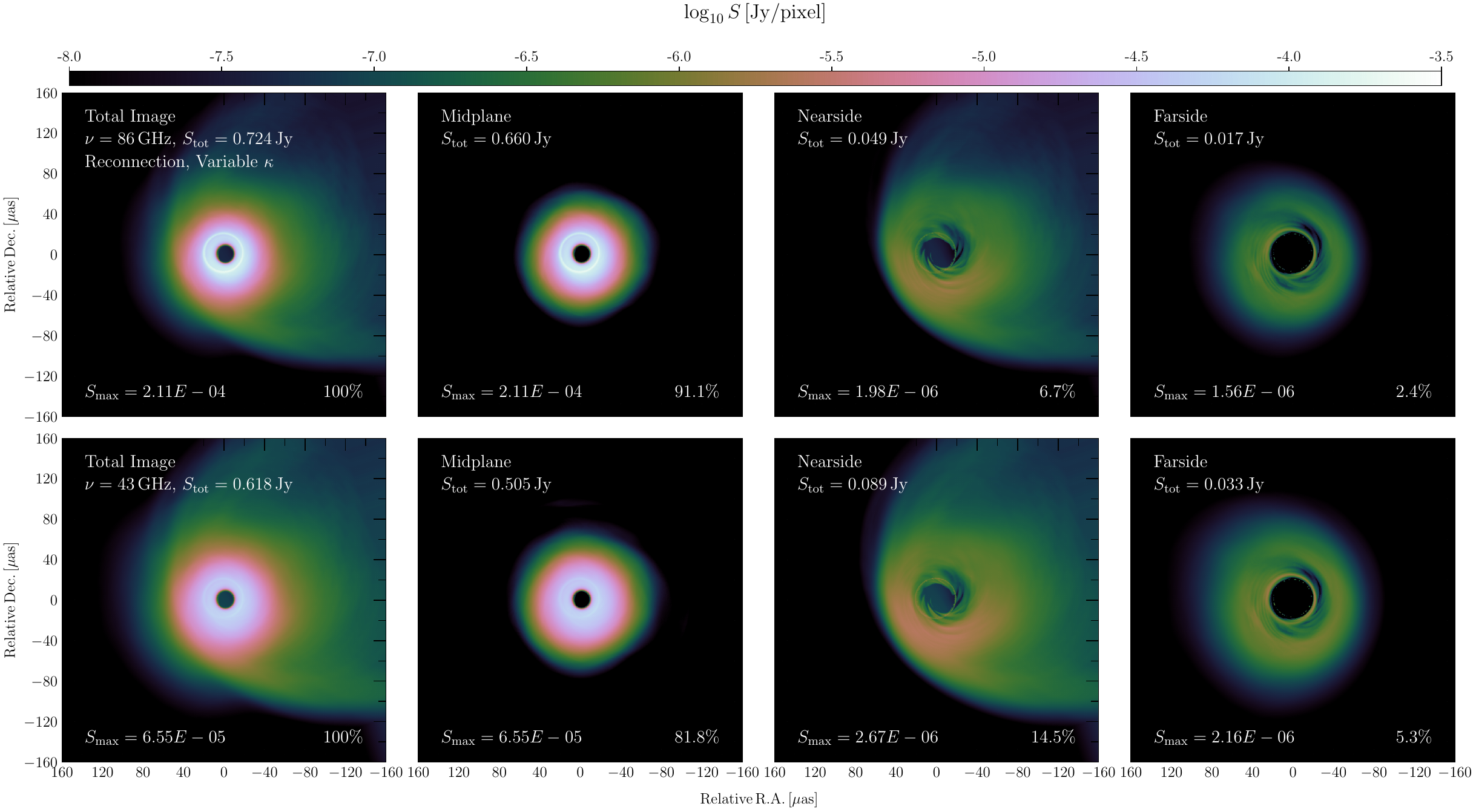}
    \caption{Time-averaged GRRT decomposed images at 86 (top) and 43~GHz (bottom) with a black hole spin of $a = 0.9375$ using the reconnection heating model with an inclination angle of $163^{\circ}$. 
    From left to right, the emissions come from every region: the whole region is depicted first, then the midplane, the nearside jet, and farside jet, respectively.
    The eDF of all of the images used is variable $\kappa$ and the time-averaged total flux is $0.5$ Jy at 230~GHz from $t=14\,000$ to $15\,000 \,\mathrm{M}$.}
    \label{figure5}
\end{figure*}

Decomposed images are designed to evaluate what portion of the emission coming from the different regions contributes to the total images. 
We divided the three regions, the midplane, the nearside jet, and the farside jet, into decomposed images, 
following the previous study \citep{2019ApJ...875L...5E}.
Figure~\ref{figure4} displays the changes in the ratio of the emission contributed by decomposed regions at 230~GHz in different heating models with a variable $\kappa$ eDF.
The percentage shown in the bottom right corner indicates the ratio of the emission contributed by a region relative to the total image. 
From the results, we can confirm that the diffuse extended jet structure seen in all heating models arises mostly from the emission from the nearside jet. 
The proportion of emission from the nearside jet differs slightly, ranging from $2.4 \, \%$ in the $R-\beta$ model ($R_{\mathrm{h}} = 1$) to $3.5 \, \%$ in the reconnection heating model, at 230~GHz. Strikingly, as is shown in Fig.~\ref{figure5}, the emission coming from nearside jet accounts for $6.7 \, \%$ of the total emission at 86~GHz and even reaches $14.5 \, \%$ at 43~GHz. 
It indicates that the contribution from the nearside jet emission increases at lower frequencies.

\subsection{Image Comparison}
\label{ImageComparison}
\begin{figure}
\includegraphics[width=\columnwidth]{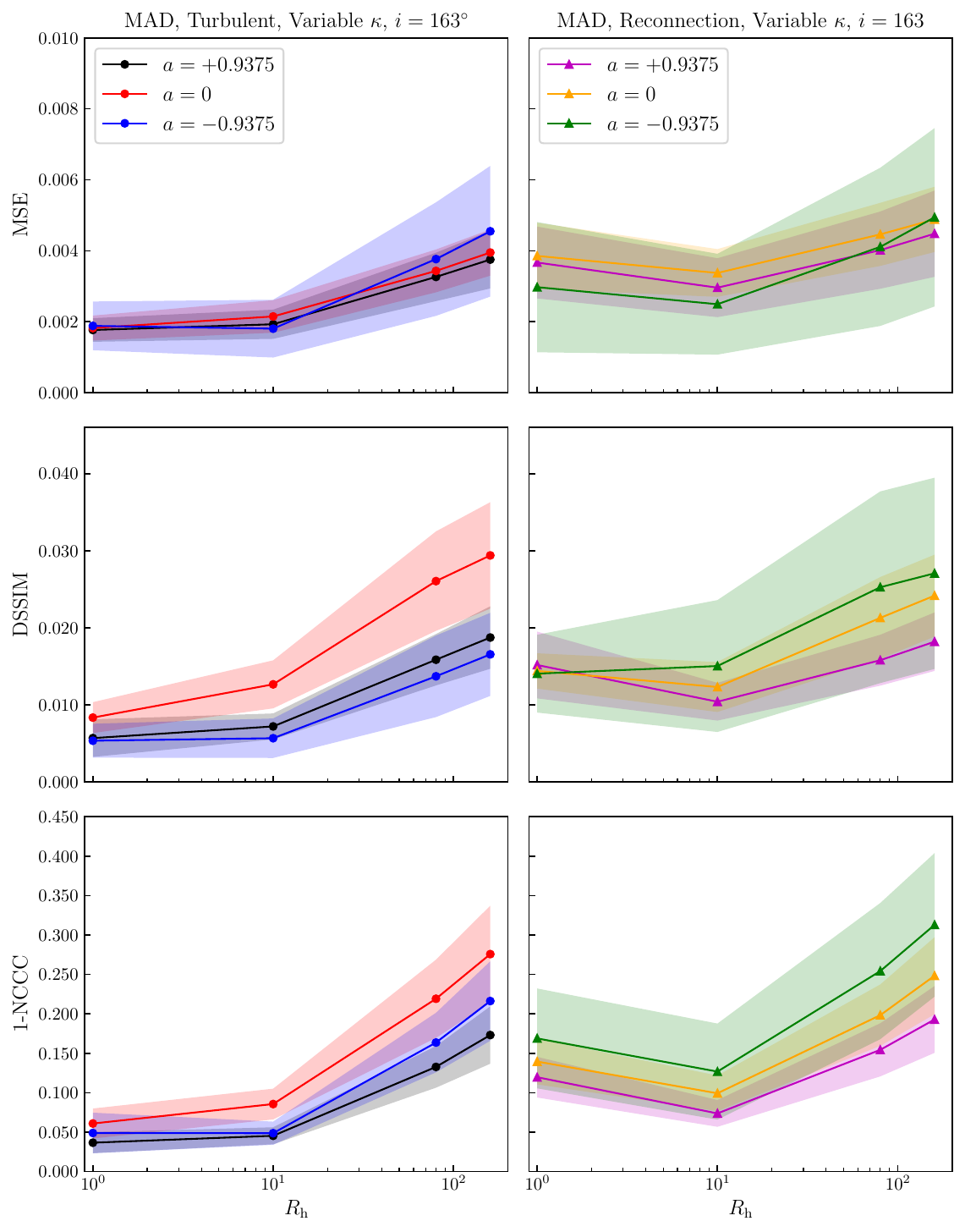}
    \caption{Image comparison between $R-\beta$ and electron heating models at different spins adopting various quantitative comparison methods with a $163^\circ$ inclination angle at 230~GHz. From top to bottom: The comparison methods used are MSE, DSSIM, and 1-NCCC, respectively.
    From left to right, the prescriptions of electron heating models are turbulent and reconnection heating, respectively.
    The applied eDF is variable $\kappa$, and the time-averaged total flux is $0.5~{\mathrm{Jy}}$ at 230~GHz. 
    The colored dots and triangles represent average values from time variation at the specific spins, and the regions shaded in the same color denote the standard deviation relative to the average values. In all of the comparison methods, the lower value indicates a close match.}
    \label{figure6}
\end{figure}

Following the previous study \citep{2021MNRAS.506..741M}, to quantitatively compare the GRRT images obtained from the $R-\beta$ models and the electron heating models, three image-comparison metrics were applied: the mean square error (MSE), the structural dissimilarity (DSSIM) \citep{2004ITIP...13..600W}, and the difference of the normalized cross-correlation coefficient (NCCC) from 1; that is, 1-NCCC \citep[][]{2018NatAs...2..585M, 2021A&A...649A.116F}. 

For the total intensity emission of two images, $I$ and $J$, the MSE was calculated pixel by pixel between two image pixels; namely, 
\begin{equation}
    \mathrm{MSE}:=\frac{\sum_{k=1}^{N}\left|I_k-J_k\right|^2}{\sum_{k=1}^{N}\left|I_k\right|^2},
\end{equation}
where $I_k$ and $J_k$, are the intensity emission at the $k$-th pixels of the two images, respectively. 
$N$ is the total number of pixels of one image, assuming that both images have the same resolution. 

Based on the human visual perception metric, the DSSIM, also known as the structural similarity index (SSIM), was calculated \citep{2004ITIP...13..600W}. The relation between them is DSSIM $=1 / \mid$ SSIM $\mid - 1$ and SSIM can be calculated as
\begin{equation}
    \operatorname{SSIM}(I, J):=\left(\frac{2 \mu_I \mu_J}{\mu_I^2+\mu_J^2}\right)\left(\frac{2 \sigma_{I J}}{\sigma_I^2+\sigma_J^2}\right),
\end{equation}
where $\mu_I:=\sum_{k=1}^{N} I_k / {N}, \sigma_I^2=\sum_{k=1}^{N}\left(I_k-\mu_I\right)^2 /({N}-1)$, and $\sigma_{I J}:=\sum_{k=1}^{N}\left(I_k-\mu_I\right)\left(J_k-\mu_J\right) /({N}-1)$. The NCCC was computed as
\begin{equation}
    \mathrm{NCCC}:=\frac{1}{N} \sum_{k=1}^{N} \frac{\left(I_k-\langle I\rangle\right)\left(J_k-\langle J\rangle\right)}{\Delta_I \Delta_J},
\end{equation}
where $\langle I\rangle$ and $\langle J\rangle$ are average pixel values of Stokes intensity, and $\Delta_I$ and $\Delta_J$ are standard deviations of pixel values \citep[e.g.,][]{2019ApJ...875L...4E}. Hence, the similarity between the two images is given by the value of NCCC, and thus $\mathrm{NCCC}=1$ denotes a strong correlation between the two images.

Image comparison between $R-\beta$ models and electron heating models at 230~GHz, with variable $\kappa$ and different black hole spins, adopting various quantitative comparison methods, are shown in Fig.~\ref{figure6}. From {$t = 14\,000$ to $15\,000 \,\mathrm{M}$}, the snapshots of a $R-\beta$ model are compared with the one with an electron heating model at the corresponding time for each $10\,\mathrm{M}$.
The lower value of the two models indicates a close match. 
The average values are represented in solid lines, while the shaded regions in the same color denote the standard deviation resulting from time variation, relative to the average values.
The results indicate that all image-comparison metrics values show an overall upward trend with the increase in $R_{\mathrm{h}}$ values, although these values are relatively small. In other words, the GRRT images of the models with the smaller $R_{\mathrm{h}}$ values are more similar to the ones of the electron heating models at 230~GHz, without the dependency on black hole spins.
This trend is the same as that seen in a previous study using thermal eDF \citep{2021MNRAS.506..741M}, while we considered variable $\kappa$, and it is also consistent with recent results of self-consistent ion-to-electron temperature from PIC turbulent simulations \citep{Meringolo2023}. In most image-comparison metrics, the values of the counter-rotating cases (blue and green) have large standard deviations. Hence, it would be difficult to find to what extent a $R-\beta$ model is comparable with the specified electron heating model. Besides, due to images at 230~GHz being already mostly optically thin, we also carried out a comparison of images at 86~GHz and 43~GHz and show the results in Appendix~\ref{AppendixC}. The aforementioned trend is also true even for lower frequencies. In addition, we also performed the same image comparison with the cases using nonthermal fixed $\kappa=3.5$ eDF. For models with fixed $\kappa=3.5$ eDF, the conclusion mentioned above does not change. Thus, this trend is independent of the eDFs. 

Apart from that, we also explored the comparison for time-averaged images at each $50\,\mathrm{M}$ to mediate the effect of small temporal structures, using variable $\kappa$ and fixed $\kappa=3.5$ eDFs, at 230 GHz, 86 GHz, and 43 GHz. The results also support the aforementioned overall increase trend with increasing $R_{\rm h}$ values. The only exception is that the MSE and 1-NCCC metrics values decrease, especially for cases with higher $R_{\mathrm{h}}$ values. At the same time, DSSIM does not change significantly, compared with results using snapshot images. The possible reason may be that the location and brightness of the extended structures are more stable, while the bright emission near the photon ring dynamically changes faster. Hence, for cases with higher $R_{\mathrm{h}}$ values, the electron in the disk region is cooler, which lowers the luminosity of bright emission and makes {a transitory structure the bigger disturbance} in the region near the center, compared with lower $R_{\mathrm{h}}$ cases. Therefore, the DSSIM, which measures the similarity of the lower-flux regions, is not sensitive to time averaging due to dim extended structures being stable. For MSE and 1-NCCC, which are more weighted in the similarity of bright regions, they decrease due to the luminous instantaneous structures being offset by a time average. 
Notably, the net decrease in MSE and 1-NCCC in cases with higher $R_{\mathrm{h}}$ values {is} larger, since such an instantaneous structure is mitigated proportionally by a longer average time, which hints that the net values, if averaged for a longer time, are reduced more in cases with higher previous MSE or 1-NCCC values.


\subsection{Spectral energy distribution}
\begin{figure}
\includegraphics[width=\columnwidth]{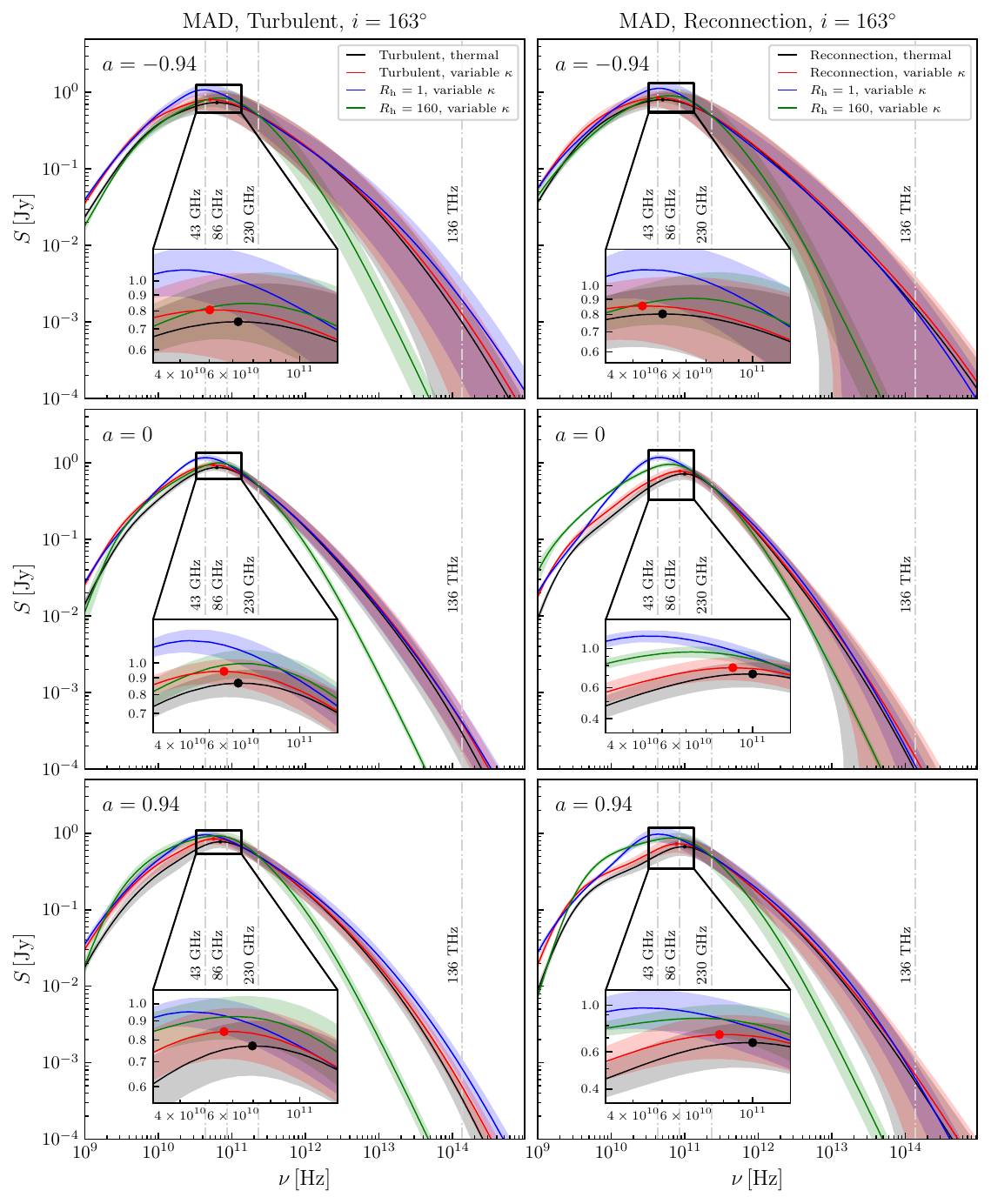}
    \caption{Spectral energy distribution curves with different eDFs and different spins in turbulent heating (left) and reconnection heating (right) models. 
    From top to bottom: Black hole spins are $-0.9375$, $0$, and $0.9375$, respectively. 
    The black curves adopt the thermal eDF, while the red curves employ variable $\kappa$ eDF.
    For the comparison, variable $\kappa$ eDF using the $R-\beta$ model with $R_{\mathrm{h}} = 1$ and $160$ are presented as blue and green curves, respectively. The solid curves represent average values and the shaded regions denote the standard deviation relative to the average values. The dash-dotted vertical lines correspond to 43~GHz, 86~GHz, 230~GHz, and 136 THz. 
    }
    \label{figure7}
\end{figure}

To investigate how flux varies with frequencies in different heating prescriptions, the {SEDs} with different electron heating models and distinct spins are shown in Fig.~\ref{figure7}. 
The solid curves represent the average flux from $t = 14\,000$ to $t = 15\,000 \,\mathrm{M}$, and the shaded regions denote the standard deviation resulting from the time variation, relative to the average values. 
Red, blue, and green curves correspond to the variable $\kappa$ eDF cases using the electron heating prescription, the $R-\beta$  model with $R_\mathrm{h}=1$, and the $R-\beta$  model with $R_\mathrm{h}=160$, respectively. 
To exhibit the impact of nonthermal eDFs, as a reference, thermal eDF cases with electron heating prescriptions are added in Fig.~\ref{figure7} in black.

First, extracting entire features is helpful to better understand the behavior of SEDs as a whole. 
As is shown in Fig.~\ref{figure7}, all of the curves behave similarly. 
As the frequency increases, the {SED} curves initially rise, followed by a decline after reaching the turnover frequency. This turnover frequency depends on both the magnetic field strength and electron temperature.\citep{2022A&A...660A.107F}.
Since the time-averaged flux at 230~GHz was set to $0.5~{\mathrm{Jy}}$, all curves intersect at the same point.
In all black hole spin cases, the standard deviation of the reconnection heating prescriptions is larger than that of turbulent heating, and the standard deviation of the $R-\beta$ model with $R_{\mathrm{h}} = 1$ is larger than that of the $R-\beta$ model with $R_{\mathrm{h}} = 160$. 
With a given electron heating prescription, the standard deviation of counter-rotating cases is the largest among cases with different black hole spins. 
A similar behavior was seen in the previous study \citep{2021MNRAS.506..741M}.

Second, with the total flux fixed at $0.5~{\mathrm{Jy}}$ at 230~GHz, the position of a peak that reflects the transition between optically thin and thick depends on the electron heating prescriptions and eDFs.
For all black hole spin cases, the curve with variable $\kappa$ eDF (red) reaches its peak at a slightly lower frequency than the one with a thermal eDF (black). The SED using the $R-\beta$ model with $R_{\mathrm{h}} = 1$ (blue) reaches its peak at a frequency comparatively lower than that using the $R-\beta$ model with $R_{\mathrm{h}} = 160$ (green). 
Thus, the peak position is sensitive to heating prescriptions.

Third, 
at high frequencies, the curves with electron heating prescriptions (black and red) and the curves with the $R-\beta$ model with $R_{\mathrm{h}} = 1$ (blue) behave similarly but the curves with the $R-\beta$ model with $R_{\mathrm{h}} = 160$ (green) deviate considerably from the other three models.
Moreover, at high frequencies, within models with variable $\kappa$ eDF, the flux of curves with $R_{\mathrm{h}} = 1$ (blue) is slightly more luminous than that of turbulent heating but for reconnection heating the trend is opposite.
Additionally, at high frequencies, the curves of electron heating prescriptions with variable $\kappa$ eDF (red) have a marginally higher flux than those with thermal eDF (black). 
It reflects the truth that nonthermal electrons contribute more to the flux at higher frequencies \citep[e.g.,][]{2019A&A...632A...2D, 2022NatAs...6..103C, 2022A&A...660A.107F}.

At low frequencies, the standard deviation becomes small due to lower time variability, and the variable $\kappa$ eDF cases have a higher flux than those in the thermal eDF. 
It becomes possible to discern the electron heating models with variable $\kappa$ eDF and those with thermal eDF through their SEDs.

In general, the contribution from the extended structure becomes larger in the emission at lower frequencies. In our studies, the FoV is limited to $\pm 160~\mathrm{\mu as}$. Thus, we may underestimate the total flux in our SED at lower frequencies.

\subsection{Time Variability}
\begin{figure*}
    \centering
    \includegraphics[width=0.45\linewidth]{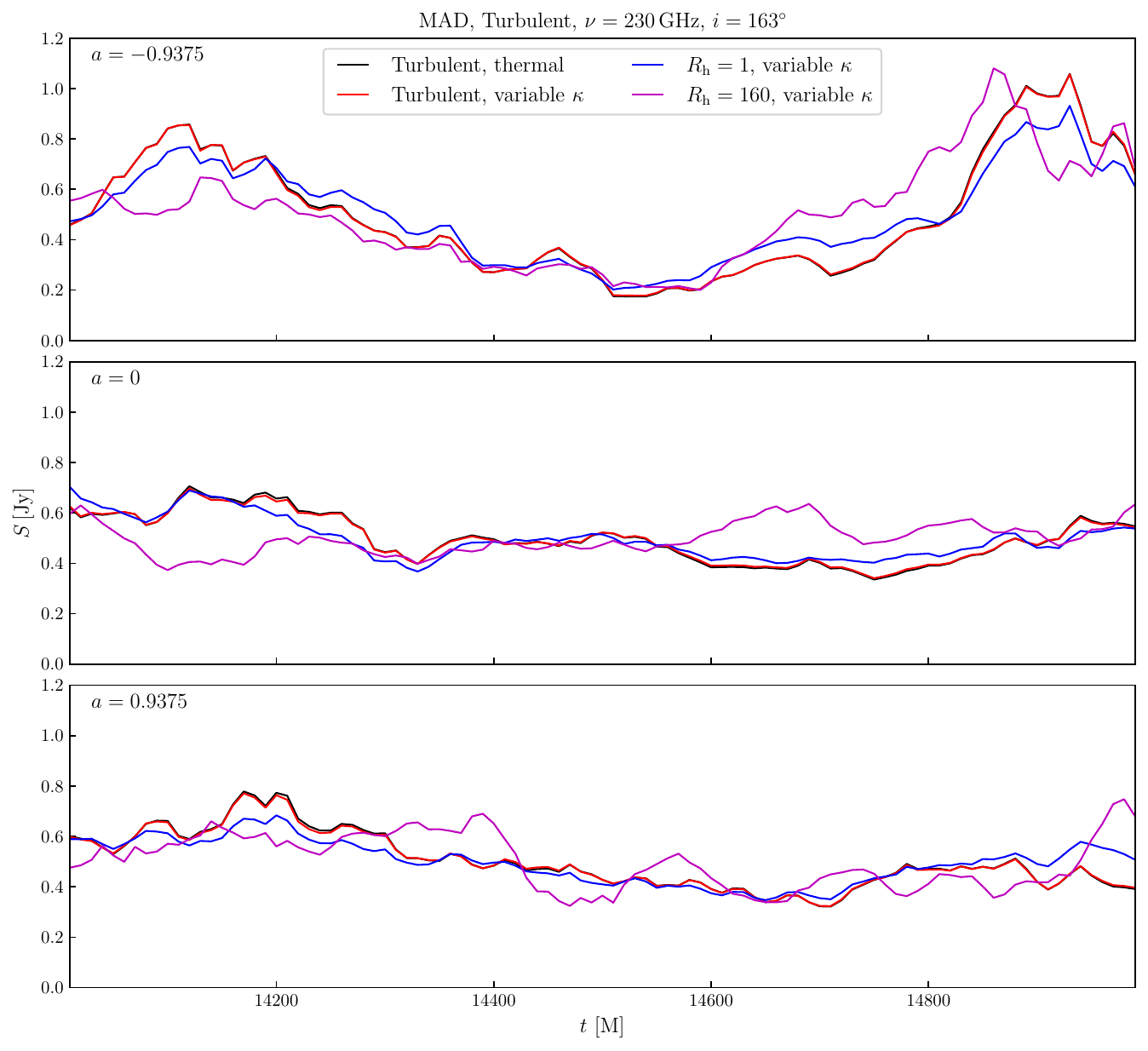}
    \includegraphics[width=0.45\linewidth]{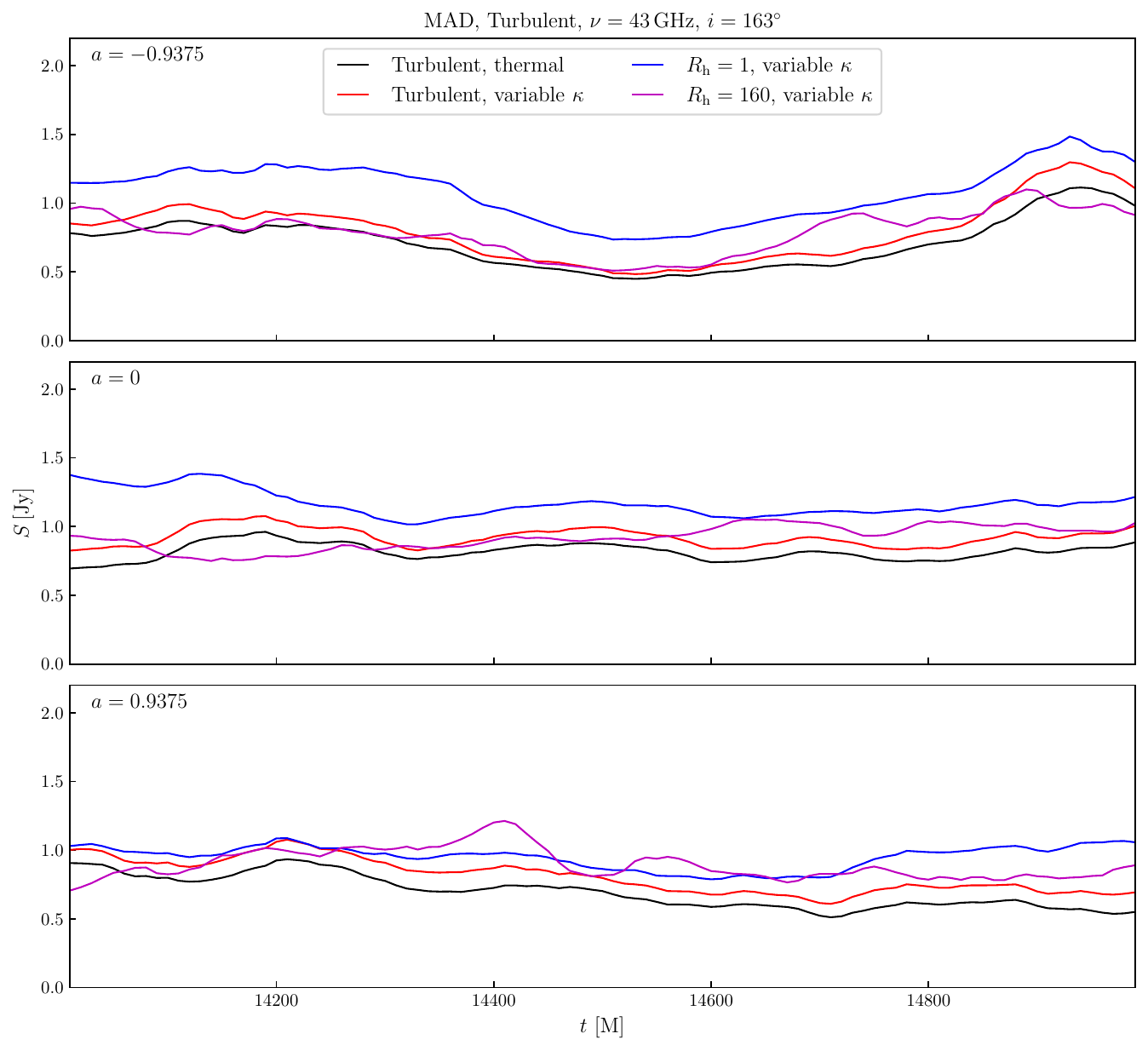}
    \caption{Light curves of flux at 230\,GHz (left) and 43\,GHz (right) with a $163^\circ$ inclination angle. Curves are plotted using $a = -0.9375$ (top), $0$ (middle), and $0.9375$ (bottom). The different colors correspond to the different heating prescriptions and eDFs: the turbulent heating model with thermal eDF (black), the turbulent heating model with variable $\kappa$ eDf (red), the $R-\beta$ model with $R_{\mathrm{h}} = 1$ in variable $\kappa$ eDF (blue), and the $R-\beta$ model with $R_{\mathrm{h}} = 160$ in variable $\kappa$ (magenta).}
    \label{figure8}
\end{figure*}
\begin{figure*}
\centering
\includegraphics[width=0.9\linewidth]{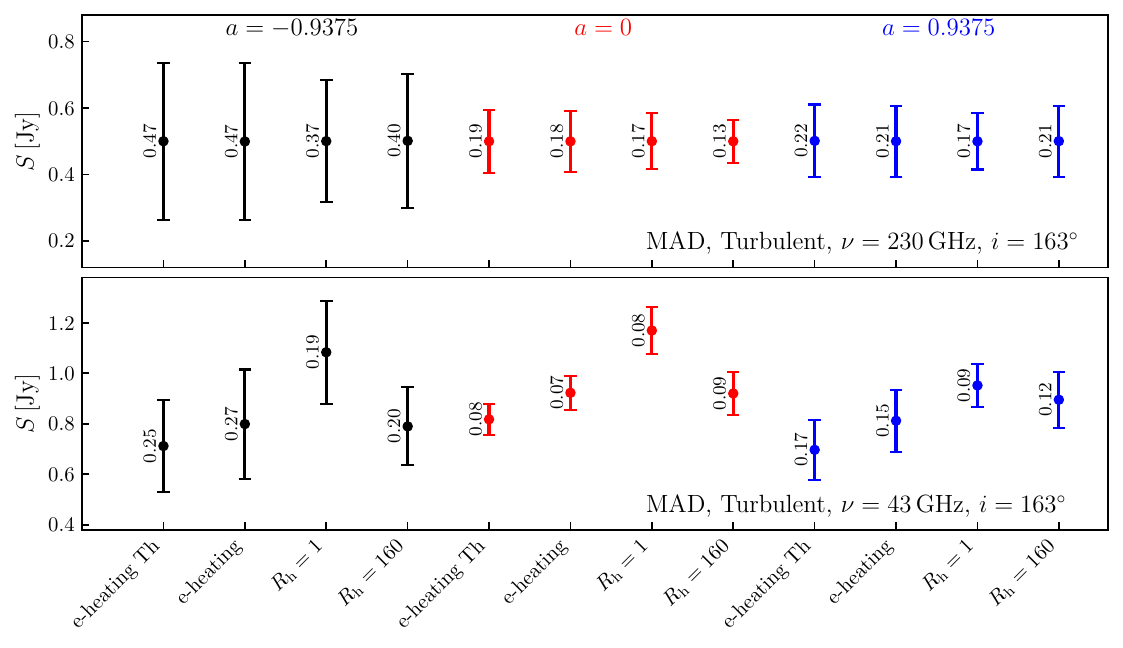}
    \caption{Total flux variation in the turbulent heating model at 230~GHz (top) and 43~GHz (bottom). The different colors correspond to the different spins: $a = -0.9375$ (black), $0$ (red), and $0.9375$ (blue). The bottom labels denote the different heating prescriptions and eDFs: the turbulent heating model with thermal eDF (e-heating Th), the turbulent heating model with variable $\kappa$ eDF (e-heating), the $R-\beta$ model with $R_{\mathrm{h}} = 1$ in variable $\kappa$ eDF ($R_{\mathrm{h}} = 1$), and the $R-\beta$ model with $R_{\mathrm{h}} = 160$ in variable $\kappa$ eDF ($R_{\mathrm{h}} = 160$). The colored dots represent average values from time variation in the specific models and the error bars denote the standard deviation relative to the average values (the number beside each error bar indicates the value of standard deviation normalized by the averaged value).}
    \label{figure9}
\end{figure*}

The light curves at 230\,GHz and 43\,GHz of models with different spins, heating prescriptions, and eDFs are shown in Fig.~\ref{figure8}.
Generally, the global trend of flux variation over time is similar in both 230\,GHz and 43\,GHz. 
We also see some small variabilities in the $R-\beta$ model with $R_{\mathrm{h}}=160$.

To quantitatively exhibit different behaviors of various models at different frequencies, the standard deviation of the time variability at 230\,GHz and 43\,GHz is shown in Fig.~\ref{figure9}. At 230\,GHz, we set the time-averaged flux at $0.5~{\mathrm{Jy}}$. Thus the average value is the same in all cases. However, the time-averaged flux at 43\,GHz changes. 
In general, at 43\,GHz, $\kappa$ eDF cases have a larger flux than those of the thermal ones, and $R-\beta$ models have a higher flux than those of the electron heating models. These are also seen in the SED profile in Fig.~\ref{figure7} and Fig.~\ref{figure8} for more clarity. In all heating models and eDFs, rotating black hole cases have higher variability than nonrotating cases and counter-rotating cases are more variable than corotating cases. 
It is a similar trend to that seen in the previous study \citep{2021MNRAS.506..741M}.
We do not see a clear difference in time variability between thermal and variable $\kappa$ eDFs. This indicates that the properties of time variability do not depend on the eDFs. 

\subsection{Exclusion of magnetized region}
\label{3.6}

{Due to the density, pressure, and internal energy possibly reaching the floor value in simulations in highly magnetized regions ($\sigma \gg 1$),} 
it is necessary to choose a ceiling to exclude regions with strong magnetization \citep{2019MNRAS.486.2873C}. The conservative cut-off value $\sigma_{\mathrm{cut}}=1$ was chosen in the previous study with thermal eDF \citep{2021MNRAS.506..741M} and also in the previous results of this paper.

\begin{figure*}
    \centering
    \includegraphics[width=0.85\linewidth]{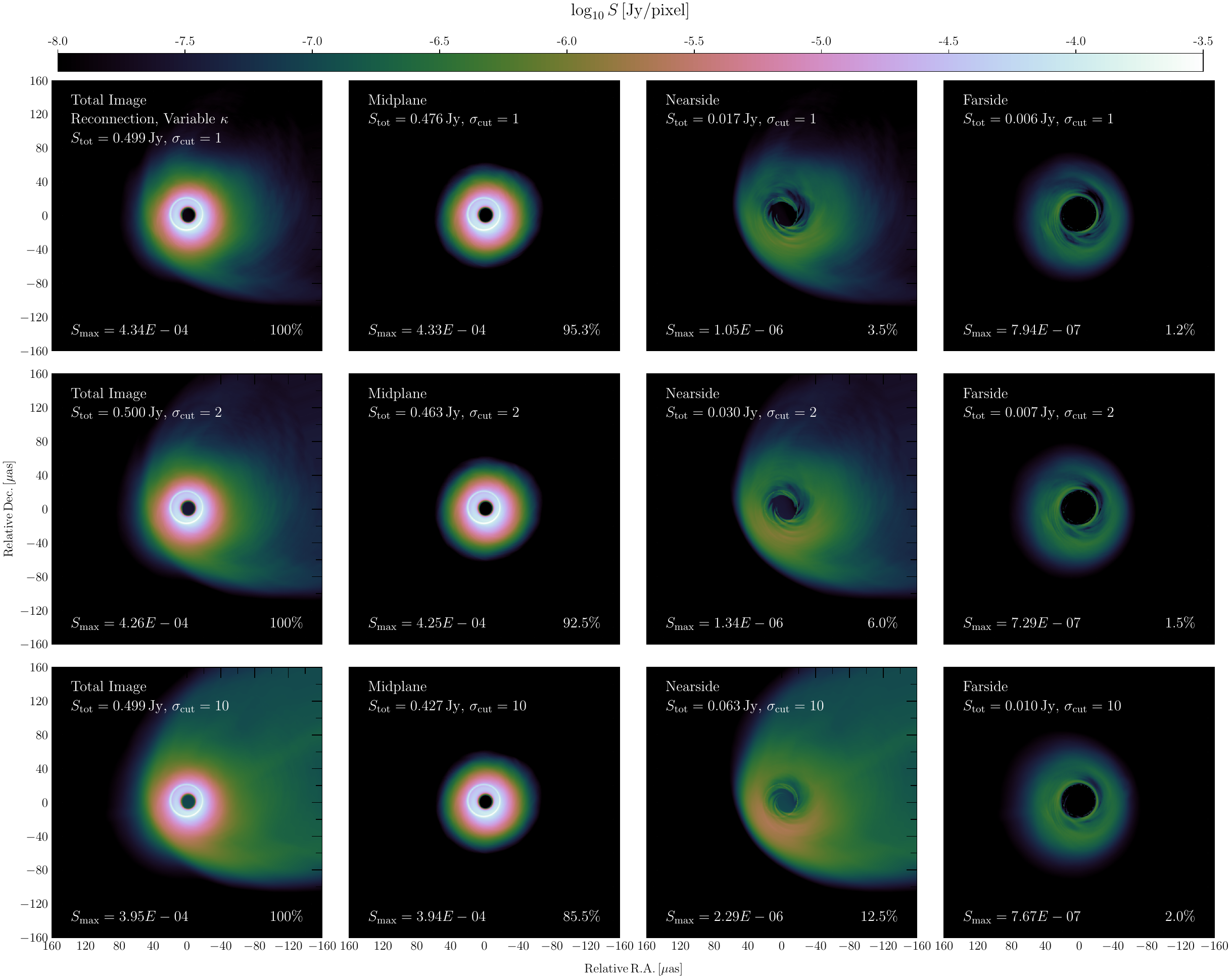}
    \caption{From $t = 14\,000$ to $t = 15\,000\,\mathrm{M}$, time-averaged MAD GRRT decomposed images with a fixed black-hole spin of $a = 0.9375$ at 230\,GHz. From top to bottom: Images using $\sigma_{\mathrm{cut}}=1$, 2, and 10, respectively. 
    From left to right, the emissions come from every region: the whole region is depicted first, then the midplane, the nearside jet, and the farside jet, respectively.
    The eDF of all of the images is variable $\kappa$, the electron heating prescription of all images is reconnection, and the time-averaged total flux of all of the images is 0.5\,Jy at 230\,GHz.}
    \label{figure10}
\end{figure*}

Here, we investigate the impacts of different $\sigma$ thresholds on images, image comparison, SEDs, and total flux variation for a MAD state with a nonthermal eDF. Figure~\ref{figure10} shows the decomposed GRRT images with various $\sigma_{\mathrm{cut}}$ at 230\,GHz with variable $\kappa$ eDF. The results show that with $\sigma_{\mathrm{cut}}$ increasing to 10,
the ratio of extended jet emission to total emission increases from $4.7 \, \%$ to $14.5 \, \%$, compared to the case of $\sigma_{\mathrm {cut}}=1$. 
In the case with $\sigma_{\mathrm{cut}}=10$, there is a faint triple-ridged jet coming from the core, although this would be affected by our floor treatment in the diffuse high magnetization regions.

\begin{figure}
    \centering
\includegraphics[width=\columnwidth]{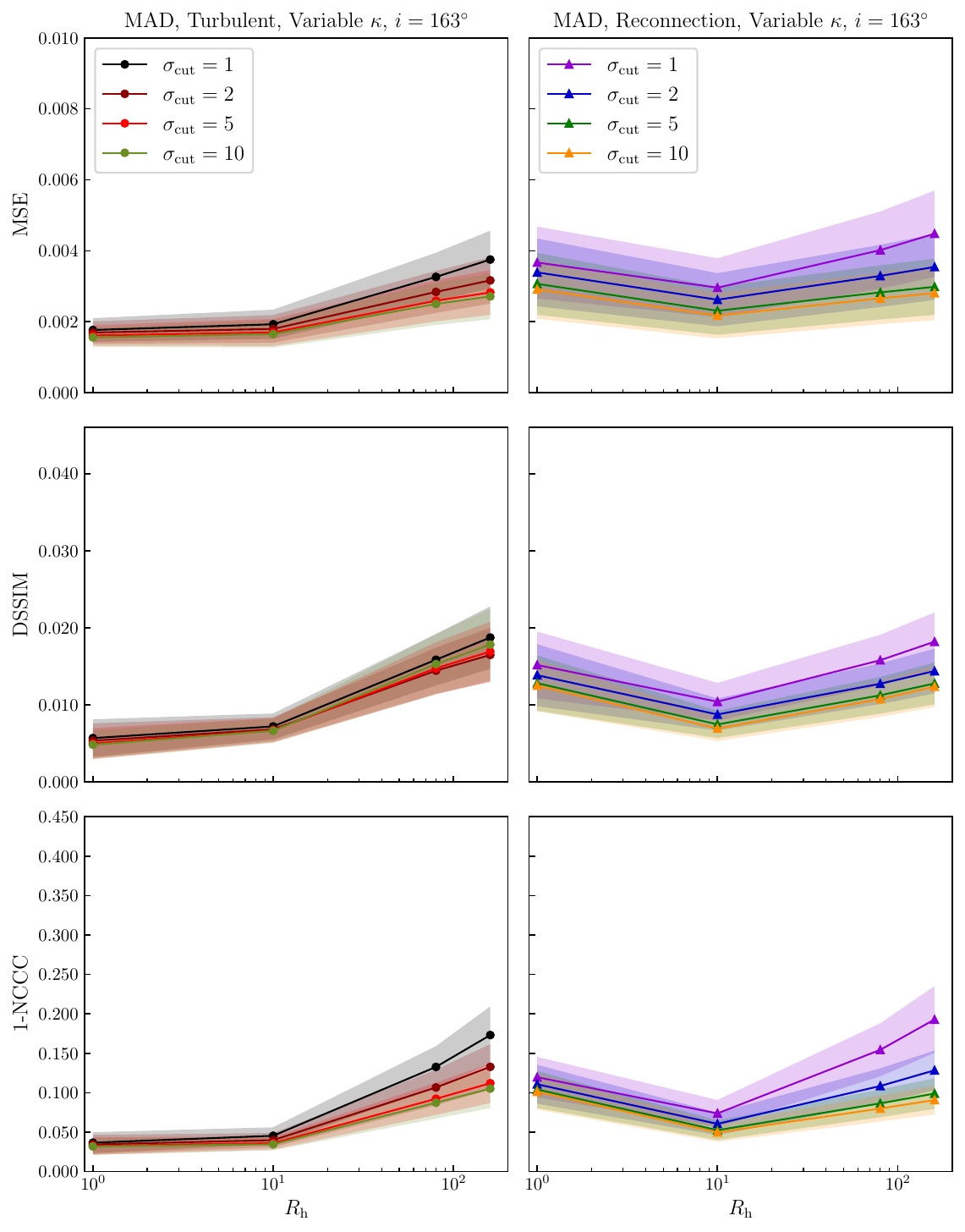}
    \caption{Same as Fig.~\ref{figure6} at 230~GHz but with various exclusion of magnetized regions: $\sigma_{\mathrm{cut}}=1$ (black and dark violet), 2 (dark red and medium blue), 5 (red and green), and 10 (olive drab and dark orange). The black hole spin in all cases is $a = 0.9375$.}
    \label{figure11}
\end{figure}

Image comparison between $R-\beta$ and electron heating models under different $\sigma_{\mathrm{cut}}$ adopting various quantitative comparison metrics are shown in Fig.~\ref{figure11}. The results indicate that the $R-\beta$ models could reproduce fairly well the images using electron heating models under different $\sigma_{\mathrm{cut}}$. With the larger $\sigma_{\mathrm{cut}}$ values, the GRRT images of a $R-\beta$ model are more similar to the ones of an electron heating model, although the increase in similarity is relatively small.

For broadband spectral energy distributions, in variable $\kappa$ eDF, the difference emerging from the choice of $\sigma_{\mathrm{cut}}$ is quite tiny up to 230\,GHz. These results are consistent with those of thermal eDF \citep{2021MNRAS.506..741M}. Thus, we skip to show the figure.

\begin{figure*}
     \centering
     \includegraphics[width=0.85\linewidth]{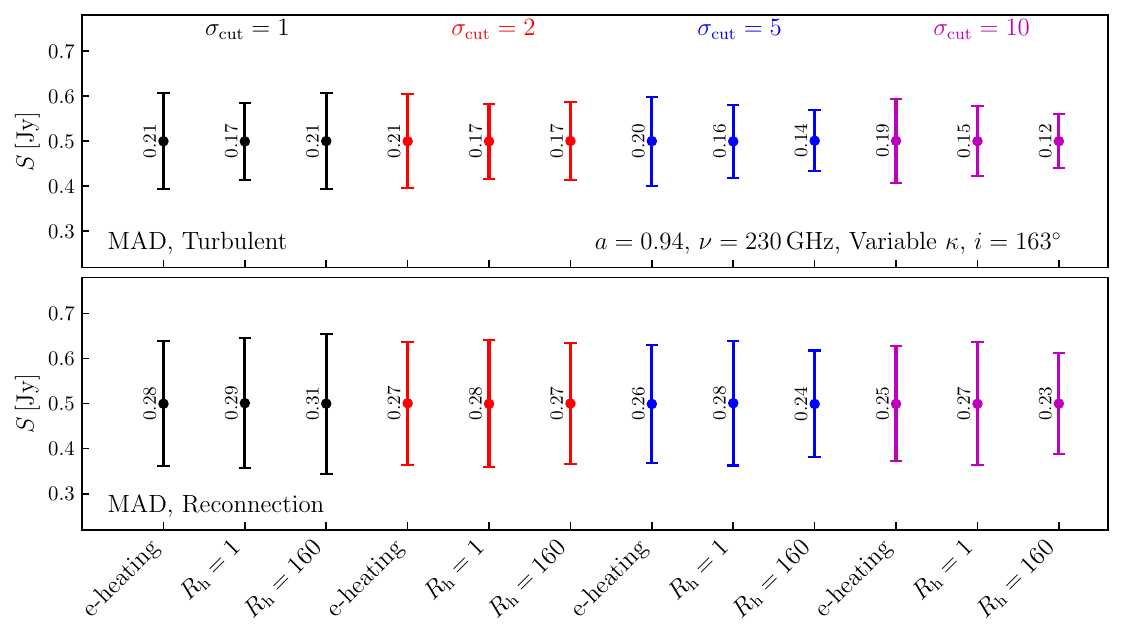}
    \caption{Total flux variation of corotating cases in the turbulent heating model (top) and reconnection model (bottom) at 230~GHz. The different colors correspond to the different exclusions of magnetized regions: $\sigma_{\mathrm{cut}}=1$ (black), 2 (red), 5 (blue), and 10 (magenta). The labels at the bottom denote the different heating prescriptions: the electron heating model (e-heating), the $R-\beta$ model with $R_{\mathrm{h}} = 1$ ($R_{\mathrm{h}} = 1$), and the $R-\beta$ model with $R_{\mathrm{h}} = 160$ ($R_{\mathrm{h}} = 160$). The colored dots represent average values from time variation in the specific models and the error bars denote the standard deviation relative to the average values (the number beside each error bar indicates the value of standard deviation normalized by the averaged value). 
    }
    \label{figure12}
\end{figure*}

The dependency of the total flux variation on $\sigma_{\mathrm{cut}}$ is shown in Fig.~\ref{figure12}. Overall, the variation in reconnection models is larger than that in turbulent heating models. Furthermore, with increasing $\sigma_{\mathrm{cut}}$, the variation in the $R-\beta$ model with $R_{\mathrm{h}} = 160$ becomes smaller, while the difference between the two temperature models and the $R-\beta$ model with $R_{\mathrm{h}} = 1$ is small. Thus, the flux variation in these models does not have a dependence on $\sigma_{\mathrm{cut}}$.


\section{Summary and discussion}
\label{discussion}

Previous research \citep{2021MNRAS.506..741M} has largely ignored the impacts of nonthermal electrons on the comparison of images between two-temperature electron heating and ordinary parameterized prescriptions. To address this issue, we adopted $\kappa$ {eDF} \citep[][]{2018A&A...612A..34D, 2019A&A...632A...2D}, which can reproduce the power law tail and which has a thermal core. Using it, we have investigated the impacts of nonthermal eDFs on images, SEDs, and time variability at 230\,GHz, 86\,GHz, and 43\,GHz with two various inclination angles ($i=163^{\circ}$ and $60^{\circ}$). We have also explored the influence of different $\mathrm{\sigma}$ thresholds on the jet morphology, shadow images, SEDs, and total flux variation.

{In a previous study \citep{2021MNRAS.506..741M}, it was found that there is not much difference between the electron heating models and the $R-\beta$ models using thermal distribution. In our study, at 230~GHz there is still no significant difference between electron heating models and the $R-\beta$ models, even if nonthermal eDF is considered. In particular, independently of heating prescriptions and the $\sigma_{\mathrm{cut}}$ values, the $R-\beta$ model with $R_{\mathrm{h}} = 1$ is the most similar to the model with electron heating in terms of the morphological structure, and the SEDs at a high frequency.}  

From our results, the extent to which the diffuse extended jet contributes to the images is related to the eDFs, electron-heating mechanisms, frequencies, and the exclusion of magnetized regions. 
Specifically, images of the variable $\kappa$ eDF are analogous to the thermal ones except for the small differences in the extended structure and the maximum flux. 
The edge-brightened jet emission structure is seen in the variable $\kappa$ eDF in corotating black hole cases in magnetic reconnection heating.
Through decomposed images, we have found that the emission coming from the nearside jet accounts for the origin of this edge-brightened jet and that it becomes more luminous at a lower frequency. 
To further investigate the contribution of the jet to total emission, the higher magnetized regions were also considered by increasing $\sigma_{\mathrm{cut}}$ from 1 to 10. 
The results show that with the growth of $\sigma_{\mathrm{cut}}$ the extended jet is more dominant. {For instance, at 230\,GHz, there will be 10~$\%$ more emission coming from the jet if including the regions with magnetization $ 1 < \sigma < 10$ when $\varepsilon=0$. Moreover, if the magnetic energy contribution to $\kappa$ eDF is considered (e.g., by setting $\varepsilon=0.5$), 10~$\%$ more emissions come from the jet contributed by electrons in magnetized regions ($ 0 \ll \sigma \leq 1$) with a radius larger than $r_{\mathrm{inj}}=10\,\mathrm{M}$.}
In addition, the energy exchange due to Coulomb coupling, anisotropic heat flow conducted along magnetic field lines, and radiative cooling considered in previous work \citep{2015MNRAS.454.1848R, 2017MNRAS.467.3604R, 2018MNRAS.478.5209C, 2019MNRAS.486.2873C} were ignored in our GRMHD simulation for simplicity.
We should note that these effects will change the electron temperature distribution and the image morphology of the shadows; for instance, diffuse extended jets may become brighter \citep{2018MNRAS.478.5209C, Dihinghia2023}.

{In particular, }from our results based on the images at lower frequencies, the extended emission contribution becomes larger.
This finding suggests that nonthermal emission at lower frequencies {such as} 43\,GHz and 86\,GHz is more prominent, and such an extended jet mission has been seen in observations \citep{2021ApJ...911L..11E}. 
{Even with the images at 230~GHz that are explained sufficiently by the Maxwell-J$\Ddot{\mathrm{u}}$ttner distribution \citep{2019ApJ...875L...5E}, the nonthermal contribution is crucial to explaining enough emission from the extended jet at other frequencies. For instance, nonthermal eDF could reproduce the emission structure of M87 at 86\,GHz \citep{2022NatAs...6..103C, 2024SciA...10N3544Y}.}
{Moreover, future} multi-frequency observations are needed to investigate the electron heating mechanism for the production of the emission of black hole shadows and extended diffuse jets.
Observation of the potential faint jet structure at 230\,GHz is crucial to imposing additional restrictions on models.
Indeed, future observations at 230\,GHz from space \citep[][]{2019BAAS...51g.256D, 2021ApJS..253....5R, 2021A&A...650A..56R} are expected to detect this faint extended structure. 
Recently, \cite{Fromm2023a} have studied the possibility of determining the electron heating model by SED and spectral index maps from the observations of future VLBI arrays.

Lastly, we note that, in this work, we only investigated the MAD model. In general, images created from the SANE model have more dependency for the choice of $R_{\mathrm h}$ value. Future research should study the SANE model to make our conclusion more robust.

\begin{acknowledgements}
    This research is supported by the National Key R\&D Program of China (2023YFE0101200), the National Natural Science Foundation of China (Grant No. 12273022), and the Shanghai Municipality orientation program of Basic Research for International Scientists (Grant No. 22JC1410600). CMF is supported by the DFG research grant "Jet physics on horizon scales and beyond" (Grant No. 443220636). ZY acknowledges support from a UKRI Stephen Hawking Fellowship.
    The simulations were performed on the Astro cluster in Tsung-Dao Lee Institute, $\pi$~2.0, and Siyuan-1 cluster in the Center for High Performance Computing at Shanghai Jiao Tong University.
    This work has made use of NASA's Astrophysics Data System (ADS). 
\end{acknowledgements}

%
   \bibliographystyle{aa} 
   \bibliography{example} 

\newpage
\begin{appendix}
\section{The effect of the magnetic energy contribution to the $\kappa$ width, $w$}
\label{AppendixA}
\begin{figure}[h]
    \begin{minipage}[t]{1\textwidth}
        \centering
        \includegraphics[width=0.8\linewidth]{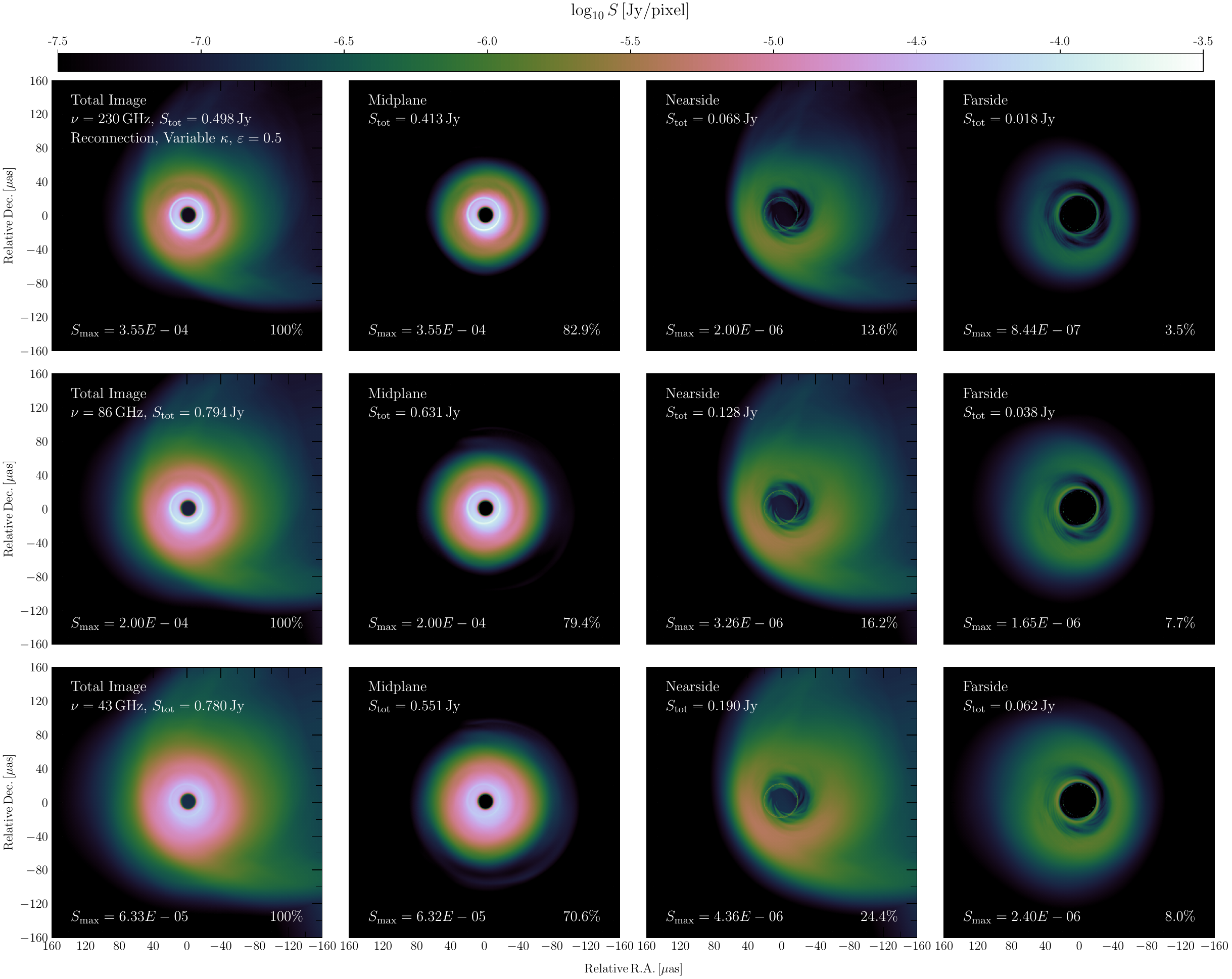}
        \caption{Time-averaged GRRT decomposed images at 230 (top), 86 (middle), and 43~GHz       (bottom) with a black hole spin of $a = 0.9375$ using the reconnection heating model with an inclination angle of $163^{\circ}$. 
        From left to right, the emissions come from every region: the whole region is depicted first, then the midplane, the nearside jet, and the farside jet, respectively.
        The eDF of all of the images used is variable $\kappa$ with $\varepsilon=0.5$ in the $\kappa$ width, $w$, and the time-averaged total flux is $0.5$ Jy at 230~GHz from {$t=14\,000$ to $15\,000 \,\mathrm{M}$.}
        }
        \label{figure13}
    \end{minipage}
\end{figure}

\begin{figure}[h]
    \begin{minipage}[t]{0.5\textwidth}
        \centering
        \includegraphics[width=1\linewidth]{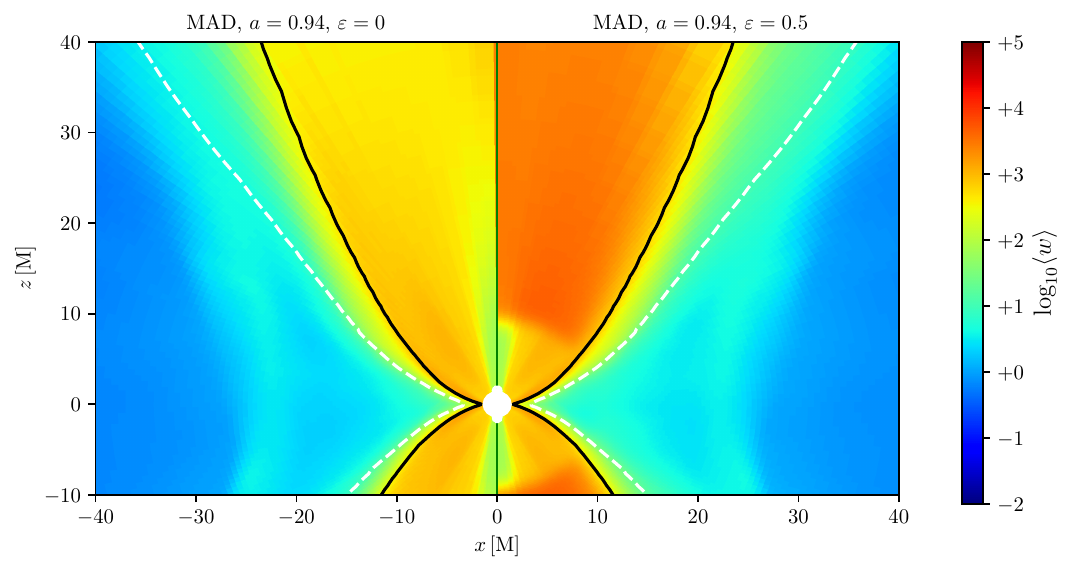}
        \caption{Time and azimuthally averaged width, $w$, of the $\kappa$ distribution on a logarithmic scale, with the reconnection heating prescription              ($a=0.94$). The left panel shows the width, $w$, with $\varepsilon=0$, and the right one shows the nonzero $\varepsilon$ model with $\varepsilon=0.5$.     The solid black and dashed white lines represent $\sigma=1$ and the             Bernoulli parameter, $-hu_{\mathrm{t}}=1.02$, respectively.
        }
        \label{figure14}
    \end{minipage}
\end{figure}

\begin{minipage}[t]{0.5\textwidth}
To investigate the effect of the magnetic energy contribution to the width, $w$, of the $\kappa$ distribution at different frequencies, we performed the GRRT calculation using variable $\kappa$ and set $\varepsilon=0.5$ in {Eq.}~\eqref{eq4}. As is shown in Fig.~\ref{figure13}, the emission originating from nearside jet accounts for 13.6\% of the total emission at 230~GHz and reaches 24.4\% at 43~GHz.
It shows that the contribution from the nearside jet emission increases at lower frequencies, which is consistent with the trend we found in Fig.~\ref{figure3} and Fig.~\ref{figure4}, in which $\varepsilon=0$. However, there is a constant enhancement of around 10\% on the nearside jet at {these three} frequencies, compared with the results using $\varepsilon=0$. {To imagine how the $\kappa$ width, $w$, of the $\kappa$ distribution in {Eq.}~\eqref{eq4} distributes around the black hole, {Figure}~\ref{figure14} shows the time and azimuthally averaged width, $w$, of the $\kappa$ distribution with $\varepsilon=0$ and $0.5$.} In the case of the width, $w$, with $\varepsilon=0$, it does not consider the contribution of magnetic energy in highly magnetized regions; that is, the jet sheath region. When considering a nonzero $\varepsilon$ case such as $\varepsilon=0.5$, {the $w$ is significantly enhanced for those regions with a radius larger than $r_{\mathrm{inj}}=10\,\mathrm{M}$ and $\sigma > 1$}, {and the $w$ is also enhanced for those magnetized regions ($ 0 \ll \sigma \leq 1$) with a radius larger than $r_{\mathrm{inj}}=10\,\mathrm{M}$}. {Therefore,} more emissions come from the jet, and the extended structure becomes more luminous. Due to the total flux being fixed, the portion of emission coming from the nearside jet increases at different frequencies.
\end{minipage}
\newpage
\newpage
\section{Decomposed images at inclination $60^{\circ}$}
\label{AppendixB}
\begin{figure}[h]
    \begin{minipage}[t]{1\textwidth}
        \centering
        \includegraphics[width=0.85\linewidth]{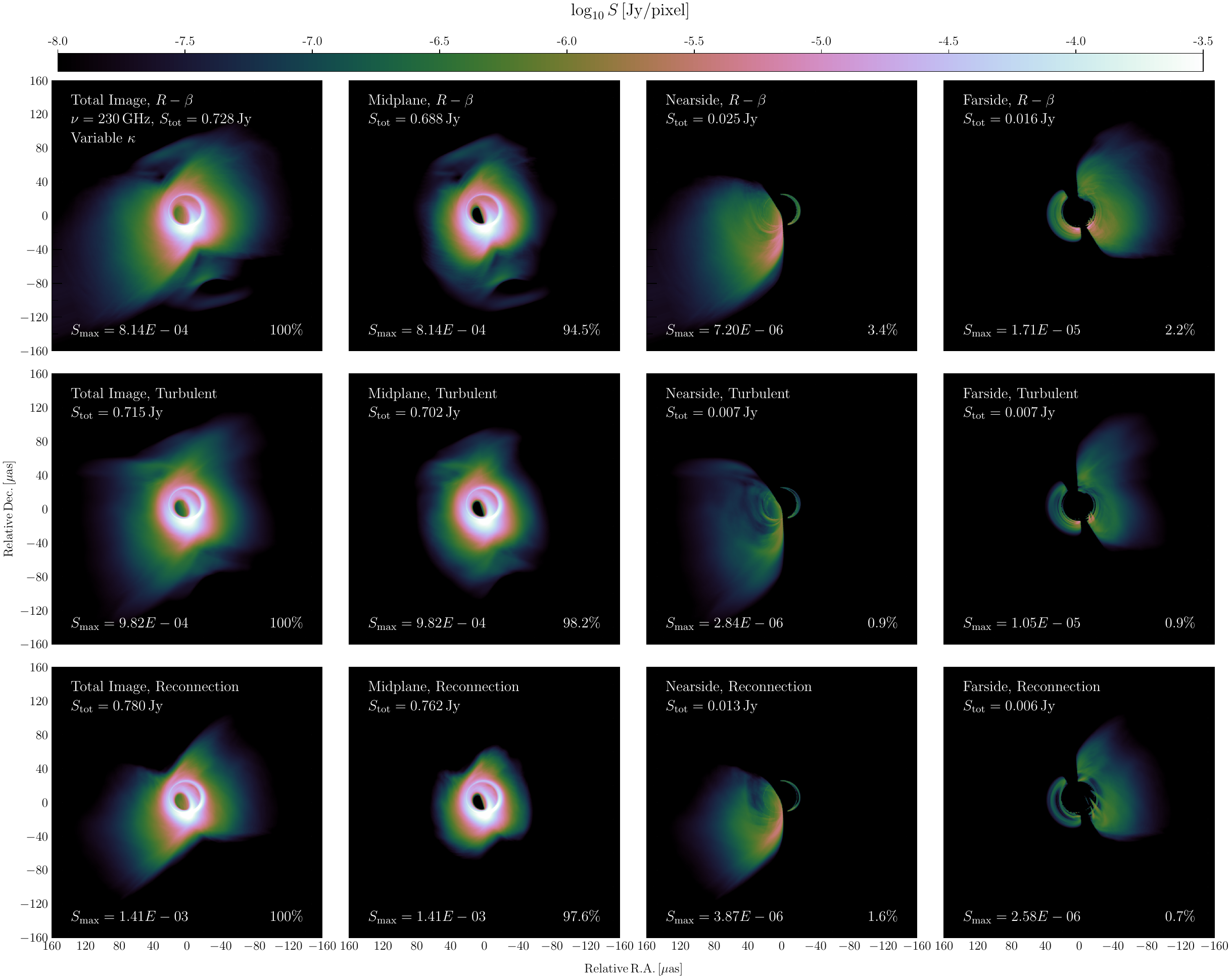}
        \caption{Same as Fig.~\ref{figure4} but with an inclination angle of                $60^{\circ}$.
            }
        \label{figure15}
    \end{minipage}
\end{figure}

\begin{minipage}[t]{0.5\textwidth}
As is shown in Fig.~\ref{figure3}, the images using reconnection heating lead to a smaller vertical extended structure at an inclination angle of $60^{\circ}$, in contrast with the ones of the turbulent heating or the $R-\beta$ prescriptions. To understand the origin of this vertically extended structure, we calculated the decomposed images following the same procedure as in Fig.~\ref{figure4}.
We should note that the accretion rates for the calculation of images at $60^{\circ}$ are the same as those at $163^{\circ}$. The results with black hole spin $a = 0.9375$ are shown in Fig.~\ref{figure15}. Because the decomposed images are calculated by the limitation of the simulation domain, the emission from the equatorial plane is omitted for nearside and farside images, which are shown as vertical discontinuous dark areas in those images. 
Clearly, the turbulent model results in a vertically extended structure, compared with the reconnection model. As is seen in the decomposed images at $163^{\circ}$ (Fig.~\ref{figure4}), emission from midplane in the turbulent heating model is more dominant than those in the reconnection heating model. 
Such a difference is enhanced as the difference of vertically extended structure in the images at $60^\circ$.
\end{minipage}
\clearpage
\section{Image comparison at 86 and 43~GHz}
\label{AppendixC}
\begin{figure}[h]
    \begin{minipage}[t]{1\textwidth}
        \centering
        \includegraphics[width=0.45\linewidth]{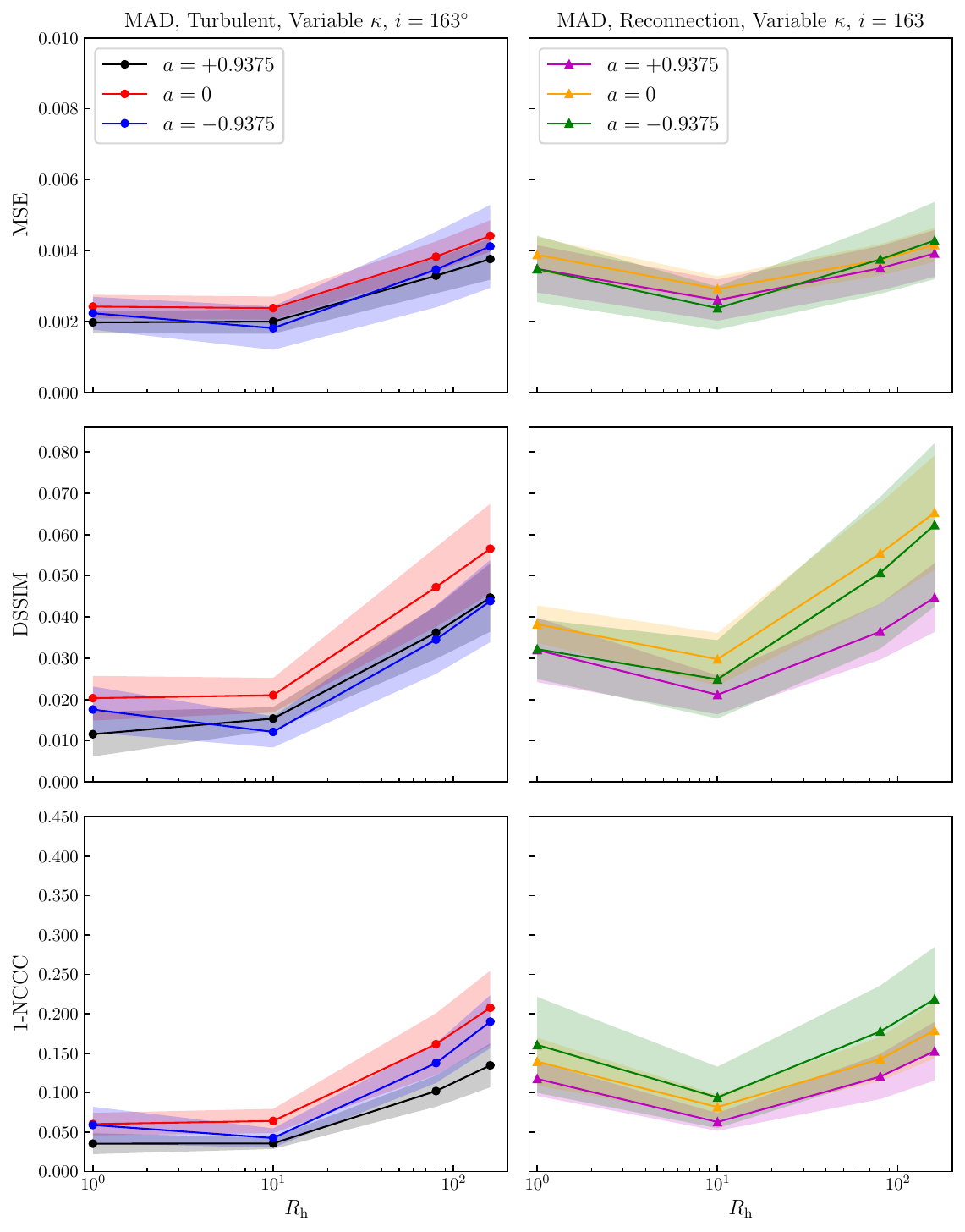}
        \includegraphics[width=0.45\linewidth]{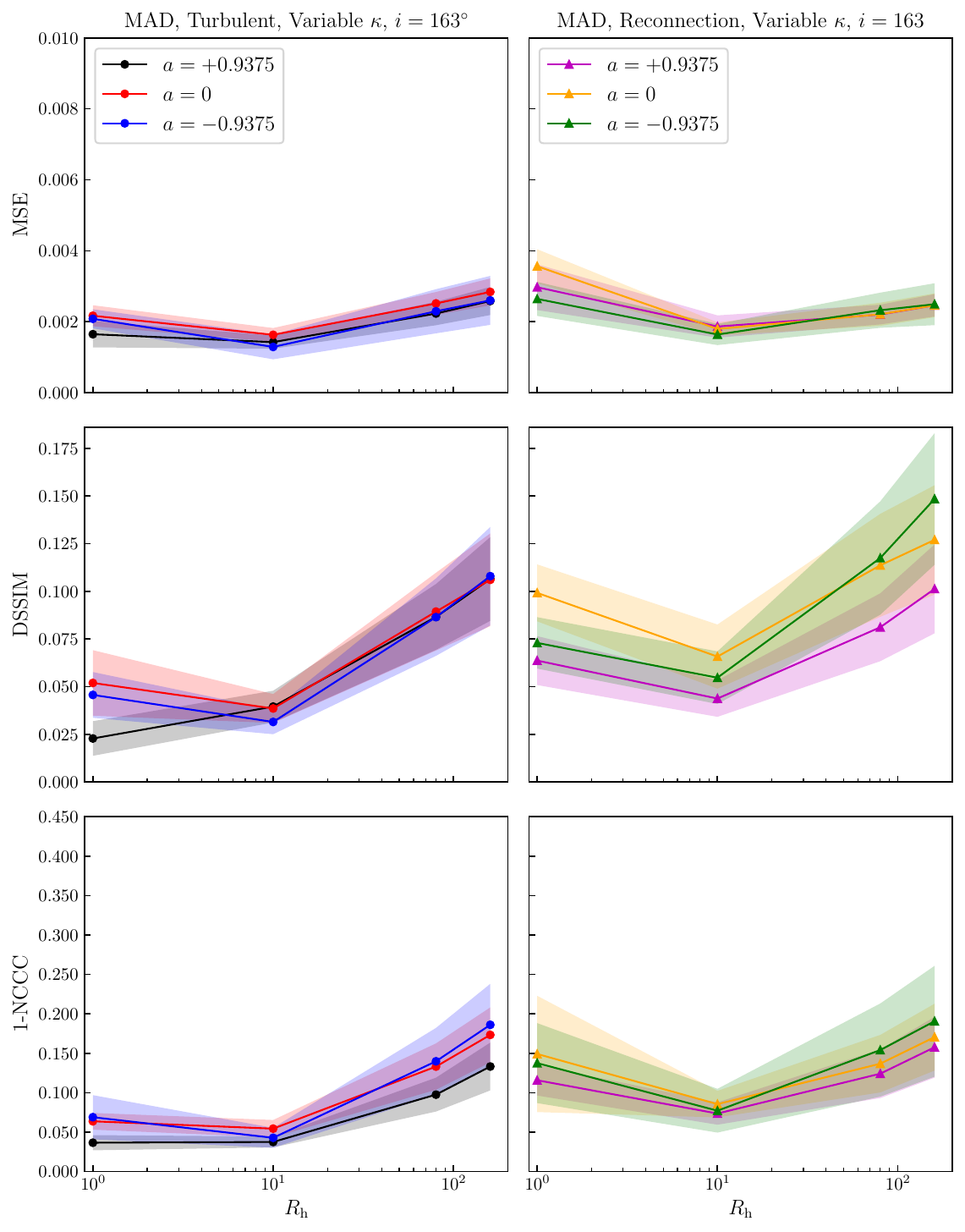}
        \caption{Same as Fig.~\ref{figure6} but at 86~GHz ({\it left}) and 43~GHz ({\it         right}).
                }
        \label{figure16}
    \end{minipage}
\end{figure}

\begin{minipage}[t]{0.5\textwidth}
Due to the images at 230~GHz being almost optically thin, it is useful to investigate how the similarity between $R-\beta$ model and two-temperature models varies with $R_{\mathrm{h}}$ values at lower frequencies. 
To do it, {we} performed a snapshot image comparison between the two-temperature electron heating model and $R-\beta$ model with different $R_{\mathrm{h}}$ at 86~GHz and 43~GHz.
Figure~\ref{figure16} provides the quantitative image comparison at 86~GHz and 43~GHz, using three different metrics described in section~\ref{ImageComparison}. Smaller values indicate more similarity between the two corresponding compared models. The results also indicate that all of the image-comparison metric values exhibit an overall increasing trend with the increase in $R_{\mathrm{h}}$ values. In addition, the DSSIM values become higher at lower frequencies, which hints that there is more divergence of diffuse, extended structures between $R-\beta$ models and electron heating models at lower frequencies.
\end{minipage}
\clearpage
\section{The equilibrium radius of general relativistic magnetohydrodynamic simulations}
\begin{figure}[h]
    \begin{minipage}[t]{1\textwidth}
        \centering
        \includegraphics[width=0.85\linewidth]{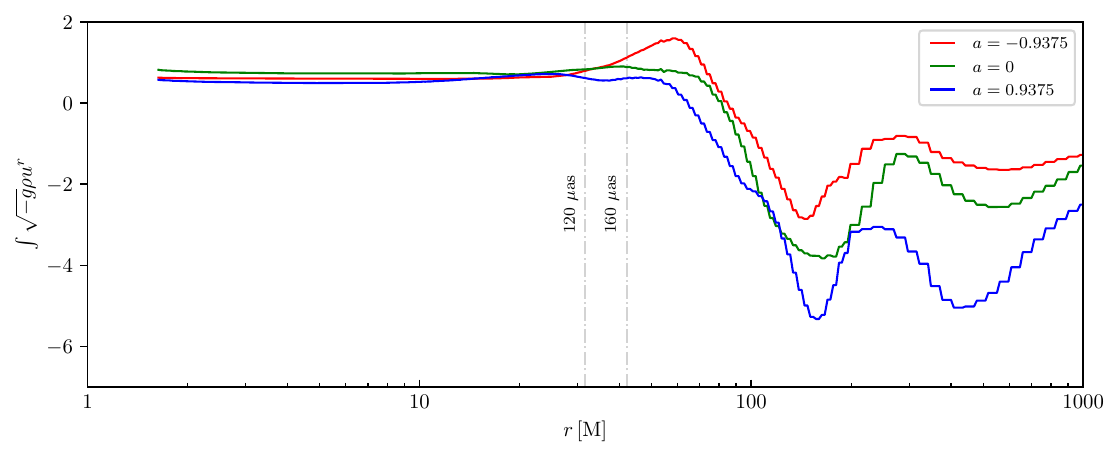}
        \caption{Time-averaged vertically integrated mass flux with black hole spins of $a = -0.9375$ (red), $0$ (green), and $0.9375$ (blue) in the turbulent heating prescription. The dash-dotted vertical lines correspond to the apparent radius of 120 and 160 $\mathrm{\mu as}$ FoV in the GRRT images.}
        \label{figure17}
    \end{minipage}
\end{figure}

\begin{minipage}[t]{0.5\textwidth}
As was mentioned in Sec.~\ref{2.1}, the GRMHD simulations reach a quasi-stationary state at $t = 15\,000 \, \mathrm{M}$, and thereafter we discussed the impacts of nonthermal distribution on GRRT images based on these GRMHD data. 
Typically, a longer simulation time provides sufficient time for the accretion flows to reach the equilibrium state with a larger radius.
To confirm that the nonthermal extended emission stems from the relaxation region rather than the regions with very large radii, where there the equilibrium state at $t = 15\,000 \, \mathrm{M}$ has still not yet been reached, it is necessary to provide convincing proof of the relaxation of these GRMHD simulations up to a desired radius at $t = 15\,000 \, \mathrm{M}$.
Figure~\ref{figure17} provides the time-averaged vertically integrated mass flux ($\int \sqrt{-g} \rho u^r$, where g is the metric determinant, $\rho$ is the fluid rest-mass density, and $u^r$ is the radial component of the 4-velocity) as a function of the radius for different spin cases using the turbulent heating prescription. The net mass flux of inflow and outflow is integrated over $\theta$ from 0 to $\pi$ for a given radius, and this value will reach a constant for those regions when it reaches the equilibrium. In the nonrotating and corotating black hole cases, even at $t = 15\,000 \, \mathrm{M}$, Figure~\ref{figure17} shows that they have reached an equilibrium at the radius greater than the current FoV of the GRRT images ($\sim 40\,\mathrm{M}$). In counter-rotating cases, since the ISCO position, $R_{\mathrm{ISCO}} \approx 8.8 \, \mathrm{M}$, is far from the black hole horizon, it takes more time to reach the equilibrium state for a given radius. However, for current GRRT images, almost all emissions are coming within 120 $\mathrm{\mu as}$, which corresponds to $31.6\,\mathrm{M}$, and within this radius all cases reach an equilibrium state. For cases using the reconnection heating prescription, the discussion of equilibrium radius is similar, and we only discuss the turbulent cases here for simplicity.
\label{AppendixD}
\end{minipage}
\end{appendix}

\end{document}